\documentclass[aps,prd,twocolumn,showpacs,superscriptaddress,groupedaddress, nofootinbib]{revtex4} 

\usepackage{graphicx}
\usepackage{dcolumn}
\usepackage{bm}
\usepackage{color}

\usepackage[colorlinks=true]{hyperref}
\hypersetup{citecolor = blue}

\usepackage{newtxtext,newtxmath}
\usepackage{mathrsfs}
\usepackage[utf8]{inputenc}
\usepackage{graphicx}                           
\usepackage{amsmath}	
\usepackage{pdflscape}	
\usepackage[T1]{fontenc}
\usepackage{ae,aecompl}
\usepackage{ulem}

\begin{document}

\preprint{APS/123-QED}

\title{Cosmological constraints from neighbor-density-weighted marked correlation functions}%

 \author{Xu Xiao}
 \affiliation{School of Physics and Astronomy, Sun Yat-Sen University, Zhuhai 519082, China}

\author{Zhao Chen}
\affiliation{Tsung-Dao Lee Institute, Shanghai Jiao Tong University, Shanghai 200240, China}
\affiliation{Department of Astronomy, School of Physics and Astronomy, Shanghai Jiao Tong University, Shanghai 200240, China}
\affiliation{State Key Laboratory of Dark Matter Physics, School of Physics and Astronomy, Shanghai Jiao Tong University, Shanghai 200240, China}
\affiliation{Key Laboratory for Particle Astrophysics and Cosmology (MOE)/Shanghai Key Laboratory for Particle Physics and Cosmology, Shanghai 200240, China}

\author{Yu Yu}
\affiliation{Department of Astronomy, School of Physics and Astronomy, Shanghai Jiao Tong University, Shanghai 200240, China}
\affiliation{State Key Laboratory of Dark Matter Physics, School of Physics and Astronomy, Shanghai Jiao Tong University, Shanghai 200240, China}
\affiliation{Key Laboratory for Particle Astrophysics and Cosmology (MOE)/Shanghai Key Laboratory for Particle Physics and Cosmology, Shanghai 200240, China}

 \author{Xiao-Dong Li}
 \email{lixiaod25@mail.sysu.edu.cn}
 \affiliation{School of Physics and Astronomy, Sun Yat-Sen University, Zhuhai 519082, China}
 \affiliation{Peng Cheng Laboratory, Shenzhen, Guangdong 518066, China}

\affiliation{CSST Science Center for the Guangdong–Hong Kong–Macau Greater Bay Area, SYSU, Zhuhai 519082, China}

\author{Yiqi Huang}
\email{yq_huang@sjtu.edu.cn}
\affiliation{School of Physics and Astronomy, Sun Yat-Sen University, Zhuhai 519082, China}
 \author{Le Zhang}
 \email{zhangle7@mail.sysu.edu.cn}
 \affiliation{School of Physics and Astronomy, Sun Yat-Sen University, Zhuhai 519082, China}
 \affiliation{Peng Cheng Laboratory, Shenzhen, Guangdong 518066, China}
 \affiliation{CSST Science Center for the Guangdong–Hong Kong–Macau Greater Bay Area, SYSU, Zhuhai 519082, China}
\date{\today}

\begin{abstract}
We investigate whether neighbor-density-weighted marked correlation functions (MCFs) can extract cosmological information beyond the standard redshift-space two-point correlation function (2PCF). Using the Kun suite of 129 $w_0w_a$CDM$+\sum m_\nu$ simulations in $1~h^{-1}{\rm Gpc}$ boxes, we construct Gaussian-process emulators for the normalized scale statistic $\widehat{W}^{\alpha}(s)$ and the angular statistic $\widehat{W}^{\alpha}_{\Delta s}(\mu)$. We perform joint analyses combining multiple mark parameters $\alpha$ and quantify the information gain using the FoM in the $\Omega_m$--$\sigma_8$ plane. Relative to the 2PCF case, three-mark combinations improve the FoM by factors of $1.7$--$2.5$, while five-mark combinations increase the gain to $1.9$--$2.4$, depending on the statistic and mark definition.  We further compare density and normalized-gradient marks, finding that they are nearly redundant for isotropic statistics but complementary for angular statistics, where their combination improves the FoM by up to $43\%$. Tests of scale range and halo selection show that the marked statistics remain robust under changes in analysis choices, with the angular statistic retaining additional cosmological information that is less sensitive to tracer selection. Our results demonstrate that MCFs substantially enhance cosmological constraints beyond the standard 2PCF and provide a robust probe for next-generation galaxy surveys.

\end{abstract}

\maketitle

\section{Introduction}
\label{sec:introduction}

Modern galaxy surveys have established large-scale structure (LSS) as one of the most powerful and precise probes of cosmic expansion and structure growth. In particular, Stage-III surveys such as 2dFGRS, WiggleZ, and SDSS have firmly established two-point clustering statistics as a standard cosmological tool~\citep{2df:Colless:2003wz,beutler20116df,blake2011wigglez,blake2011wigglezb,york2000sloan,Eisenstein:2005su,Percival:2007yw,anderson2012clustering,alam2017clustering}. The two-point correlation function (2PCF) and its Fourier counterpart, the power spectrum, are widely used because of their simplicity, clear physical interpretation, and sensitivity to both the expansion history and the growth of structure~\citep{Kaiser, Ballinger, Eisenstein_1998, Blake_2003, Seo_2003}. These statistics have been successfully applied to a range of surveys, including 2dFGRS, 6dFGS, WiggleZ, and SDSS, yielding robust cosmological constraints~\citep{2dFGRS,6dFGRS,WiggleZ2011B,WiggleZ2011c,SDSS_York,Eisenstein:2005su,Percival:2007yw,anderson2012clustering,sanchez2012clustering,sanchez2013clustering,anderson2014clustering,samushia2014clustering,ross2015clustering,beutler2016clustering,sanchez2016clustering,alam2017clustering,chuang2017clustering}.

Looking ahead, ongoing and upcoming surveys such as DESI\footnote{https://desi.lbl.gov/}, LSST\footnote{https://www.lsst.org/}, Euclid\footnote{http://sci.esa.int/euclid/}, Roman\footnote{https://roman.gsfc.nasa.gov/}, and CSST\footnote{http://nao.cas.cn/csst/} will map substantially larger cosmic volumes with higher tracer densities~\citep{desi2016,lsst2009,eucild2011,eucild2024,rst2019,gong2019csst}. These Stage-IV surveys will substantially improve statistical precision, increasing the need for summary statistics that can fully exploit the available information. However, the limitations of standard two-point statistics are well recognized. Nonlinear gravitational evolution, galaxy bias, and redshift-space distortions induce non-Gaussian features in the density field that cannot be fully captured by an unweighted two-point statistic. As a result, part of the cosmological information is inevitably lost when compressing the data into the 2PCF or power spectrum alone.

This limitation has motivated the development of alternative approaches that go beyond the Gaussian two-point level, including higher-order correlation functions, void statistics, and machine-learning-based summaries~\citep{Sabiu2016A&A,Slepian_2017,Sabiu_2019,ryden1995measuring,lavaux2012precision,Ravanbakhsh17,Mathuriya18,pan2020cosmological}. These methods aim to retain the robustness and interpretability of traditional statistics while capturing additional non-Gaussian information present in LSS.
Among these, marked correlation functions (MCFs) provide a particularly simple and flexible extension of the standard two-point framework~\citep{Beisbart:2000ja,Beisbart2002,Gottl2002,Sheth:2004vb,Sheth:2005aj,Skibba2006,White_2009,White2016,Satpathy:2019nvo,massara2020,Philcox2020}. In this approach, each tracer is assigned a mark that depends on its local environment, and clustering is quantified through a weighted two-point statistic. The standard 2PCF is recovered when all marks are unity. By construction, MCFs probe the environmental dependence of clustering, allowing one to emphasize overdense or underdense regions through appropriate choices of the mark. Previous studies using mock catalogs and SDSS data have demonstrated that such density-dependent weighting can extract additional cosmological information and tighten parameter constraints relative to the 2PCF alone~\citep{MCF_Yang,Xiao2022MCF,Lai2024MCF}.

A key obstacle in applying MCFs is the lack of accurate predictive models. Their nonlinear and environment-dependent nature precludes reliable analytic mappings from cosmological parameters to the statistic, while astrophysical and observational systematics--most notably galaxy bias--introduce additional uncertainties. These limitations motivate a simulation-based strategy, in which MCFs are measured across a suite of numerical simulations and interpolated over cosmological parameter space.
This approach is enabled by emulators that combine large simulation suites with machine-learning techniques to provide fast and accurate predictions. Such methods were pioneered for the matter power spectrum by CosmicEmu~\citep{2009ApJ...705..156H,2010ApJ...715..104H,2010ApJ...713.1322L}, and have since been extended to smaller scales, wider redshift ranges, and higher-dimensional parameter spaces~\citep{2014ApJ...780..111H,2017ApJ...847...50L}. Emulators have also been developed for a range of large-scale structure observables, including the halo mass function~\citep{bocquet2020miratitan}, galaxy power spectra~\citep{Kwan2015,Wibking_2019}, and the concentration--mass relation~\citep{Kwan_2013}, as well as in recent large-scale emulation efforts~\citep{Nishimichi_2019,kobayashi2020accurate,Kwan2023,Moran2023}.

In this work, we construct a simulation-based emulator for MCFs to use environmental clustering as a cosmological probe. Unlike the standard two-point correlation function, MCFs encode how clustering depends on local density and environmental transitions, making them sensitive to nonlinear structure formation and redshift-space anisotropies. We adopt a simulation-based inference approach~\citep{Cranmer_2020}, using forward simulations to learn the mapping between cosmological parameters and MCF statistics. This bypasses the need for an explicit analytic likelihood for marked statistics with nontrivial covariance and environmental weighting. We develop a Gaussian-process-regression (GPR) emulator for neighbor-density-weighted MCFs in redshift space, trained on the \textsc{Kun} simulation suite and validated with independent \textsc{Jiutian} simulations. The \textsc{Jiutian} simulation suite \cite{han2025jiutiansimulationscsstextragalactic} consists of a variety of simulations spanning different  resolutions, cosmological parameters \cite{yu2025kunsimulation},  neutrino \cite{universe11070212,Yu_2026}, dark matter \cite{2023MNRAS.526.3156H} and dark energy models  \cite{2019ApJ...875L..11Z}, as well as zoomed simulations of specific regions of the Universe~\cite{2024ApJ...966..236L}. A number of mock galaxy catalogs have been produced from the primary runs following different methodologies ~\cite{2024MNRAS.529.4958P, 2024MNRAS.529.4015G, tan2025semianalyticalmockgalaxycatalog, wei2025mockobservationscsstmission, Wei_2026}.

The framework is designed to quantify the information gain from combining multiple mark powers $\alpha$, test the complementarity between density-based and gradient-based marks, and assess the robustness of cosmological constraints to choices of scale range and halo selection. We focus on two reduced statistics, $\widehat{W}^{\alpha}(s)$ and $\widehat{W}^{\alpha}_{\Delta s}(\mu)$, which retain the main scale-dependent and line-of-sight environmental information in a compact data vector.

The paper is organized as follows. Section~\ref{sec:data} describes the simulation suites and tracer selection. Section~\ref{sec:methodology} details the construction of the MCF estimators, the Gaussian process emulator, and the covariance modeling. Section~\ref{sec:results} presents the cosmological constraints and robustness analyses, comparing MCFs with the standard two-point correlation function. Finally, Section~\ref{sec:conclusion} summarizes our main findings and outlines future directions.

\section{Data}
\label{sec:data}

\subsection{Kun simulation}

We use the \textsc{Kun} simulation suite as the training set for the emulator~\citep{yu2025kunsimulation}. The suite consists of 129 cosmologies spanning the $w_0w_a$CDM$+\sum m_{\nu}$ parameter space. Each realization is evolved with the \textsc{Gadget-4} $N$-body solver\footnote{https://wwwmpa.mpa-garching.mpg.de/gadget4/}~\citep{Springel2021gadget} in a periodic box of side length $1~h^{-1}{\rm Gpc}$ with $3072^3$ particles, corresponding to a particle mass of $2.87\,(\Omega_{cb}/0.3)\times10^9~h^{-1}M_{\odot}$. The \textsc{Kun} suite is part of the \textsc{Jiutian} simulation program developed for CSST science~\citep{han2025jiutiansimulationscsstextragalactic,gong2019csst}.

The cosmological parameter space spans baryon density $\Omega_b$, matter density $\Omega_{m}$, scalar spectral index $n_s$, Hubble constant $H_0$, primordial fluctuation amplitude $A_s$, dark energy equation-of-state parameters $(w_0, w_a)$, and the summed neutrino mass $\sum m_\nu$, with ranges
$\Omega_b \in [0.04, 0.06]$,
$\Omega_{m} \in [0.24, 0.40]$,
$n_s \in [0.92, 1.00]$,
$H_0 \in [60, 80]~{\rm km\,s^{-1}\,Mpc^{-1}}$,
$A_s \in [1.70, 2.50]\times10^{-9}$,
$w_0 \in [-1.30, -0.70]$,
$w_a \in [-0.50, 0.50]$, and
$\sum m_{\nu} \in [0.00, 0.30]~{\rm eV}$.

\begin{figure*}[htpb]
\centering
\includegraphics[scale=0.95]{"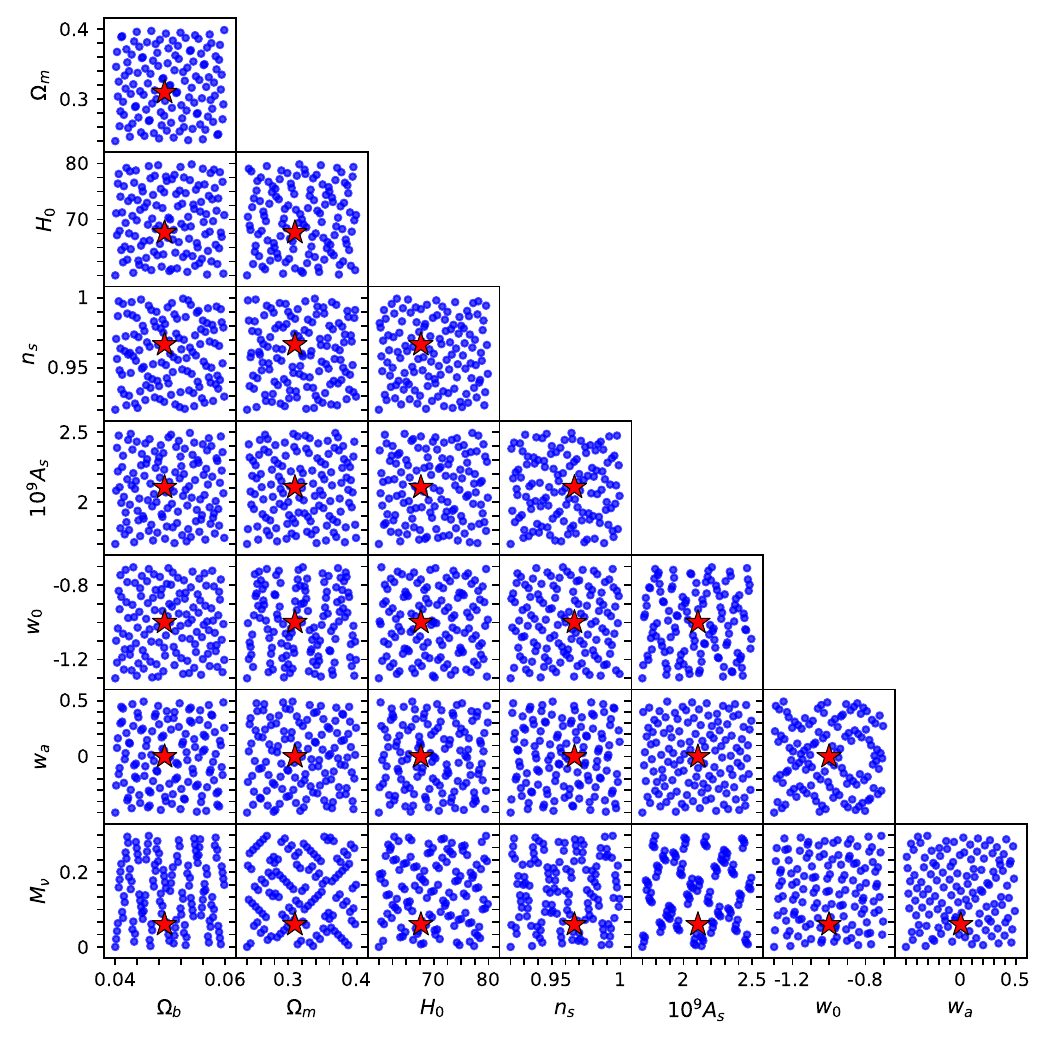"}
\caption{Sampling of the 129 \textsc{Kun} cosmologies used to train the emulator. See also Figure 2 of \citet{yu2025kunsimulation}. Blue points show the 128 Sobol-sequence cosmologies, and the red star marks the fiducial \textit{Planck} 2018 model. The broad parameter coverage ensures that the emulator operates primarily in the interpolation regime, thereby improving accuracy in the likelihood analysis.}
\label{fig:param_space}
\end{figure*}

The non-fiducial cosmologies are generated using a Sobol sequence~\citep{sobol1967distribution}, which provides a quasi-random, space-filling sampling of the high-dimensional parameter space for emulator training. Figure~\ref{fig:param_space} illustrates this distribution: the 128 Sobol-sampled cosmologies (blue points) uniformly cover the parameter volume, while the fiducial \textit{Planck} 2018 model~\citep{planck2018} is shown as a red star. This homogeneous coverage minimizes clustering and voids in parameter space, thereby improving interpolation accuracy and reducing extrapolation in the likelihood analysis. Each cosmology is realized once; to suppress large-scale sample variance, the initial conditions adopt the fixed-amplitude method~\citep{2016MNRAS.462L...1A}. Halo and subhalo catalogs are identified using the \textsc{Rockstar} algorithm~\citep{Behroozi2012rockstar}, which are used to construct the tracer samples in this work.

\subsection{Jiutian simulation}

We use the primary \textsc{Jiutian} simulation both for covariance estimation and as an independent validation dataset~\citep{han2025jiutiansimulationscsstextragalactic}. It adopts a Planck 2018 $\Lambda$CDM cosmology with parameters $\Omega_m=0.3111$, $\Omega_{\Lambda}=0.6889$, $\Omega_b=0.049$, $n_s=0.9665$, $\sigma_8=0.8102$, and $H_0=67.66~{\rm km\,s^{-1}\,Mpc^{-1}}$~\citep{planck2018}. The simulation evolves $6144^3$ particles in a periodic box of side length $2~h^{-1}{\rm Gpc}$. Halo and subhalo catalogs are identified using FoF and \textsc{Subfind}~\citep{fof,Springel_2001subfind}.

To construct tracer samples, we impose a fixed number density $\bar n=10^{-3}~(h^{-1}{\rm Mpc})^{-3}$ in all simulations by ranking halos and subhalos by mass and selecting the most massive objects. This corresponds to a characteristic mass threshold $M_{\rm cut}\simeq3.85\times10^{12}~h^{-1}M_{\odot}$ for both \textsc{Kun} and \textsc{Jiutian}. By fixing $\bar n$, we control for variations in tracer abundance across cosmologies, ensuring that differences in the measured statistics arise mainly from clustering and environmental effects. The adopted number density is comparable to that expected in current and forthcoming spectroscopic galaxy surveys.

Redshift-space positions are constructed under the distant-observer approximation, taking the simulation $z$-axis as the line of sight:
\begin{equation}
\bm{s} = \bm{r} + \frac{\bm{v}\cdot\hat{\bm{z}}}{aH(a)}\,\hat{\bm{z}},
\label{eq:rsd}
\end{equation}
where $\bm{r}$ and $\bm{s}$ denote the real- and redshift-space positions, $\bm{v}$ is the peculiar velocity, $a$ is the scale factor, and $H(a)$ is the Hubble parameter.

\section{Methodology}
\label{sec:methodology}
\subsection{Construction of marked correlation functions (MCFs) }

The purpose of MCFs is to supplement the ordinary two-point clustering signal with information about the local environment of each tracer. Instead of treating all halos equally, we assign each halo a mark that depends on its surrounding density field. Pair counts are then weighted by these marks, allowing the statistic to probe how clustering changes in dense regions, underdense regions, or across environmental transitions.

We first define the local environment using an adaptive nearest-neighbor density,
\begin{equation}
\rho_{n_{\rm NB}}(\mathbf{r})
=
\sum_{i=1}^{n_{\rm NB}}
W_k(|\mathbf{r}-\mathbf{r}_i|,h_W)\,,
\label{eq:rho}
\end{equation}
where $W_k$ is a third-order B-spline kernel with compact support inside $2h_W$ \citep{Lucy1977zz,Gingold1977smooth}. The smoothing length $h_W$ is chosen adaptively so that the kernel encloses the $n_{\rm NB}$ nearest neighbors around position $\mathbf{r}$. In this work we use $n_{\rm NB}=30$, which gives $h_W = 8.3 \pm 3.7~h^{-1}{\rm Mpc}$. The scale ($2h_W$) is thus close to the minimum pair separation used in the correlation measurement, $s_{\min}=15~h^{-1}{\rm Mpc}$. Therefore, the marks characterize quasi-nonlinear local environments, while the pair statistic is measured on larger clustering scales.

To capture not only dense regions but also spatial transitions between environments, we additionally compute the local density gradient,
\begin{equation}
\nabla \rho_{n_{\rm NB}}(\mathbf{r})
=
\sum_{i=1}^{n_{\rm NB}}
\nabla W_k(|\mathbf{r}-\mathbf{r}_i|,h_W).
\label{eq:nablarho}
\end{equation}
The gradient is useful because cosmological information may not only reside in the density amplitude itself, but also in how rapidly the environment changes around halos.

\subsubsection{Choice of marks}

We use two environmental marks in the likelihood analysis. The first is the adaptive local density itself,
\begin{equation}
w = \rho_{n_{\rm NB}}.
\label{eq:WeightRho}
\end{equation}
This mark directly upweights halos in dense environments. It is therefore sensitive to the clustering of halos in compact overdense structures.

The second mark is the normalized density gradient,
\begin{equation}
w =
\frac{|\nabla \rho_{n_{\rm NB}}|}{\rho_{n_{\rm NB}}}.
\label{eq:Weightdlnrhodr}
\end{equation}
This quantity measures the fractional change of the local density. Compared with the unnormalized gradient $|\nabla\rho_{n_{\rm NB}}|$, it reduces the direct dependence on the density amplitude and is designed to highlight relative environmental transitions.

\begin{figure*}[htpb]
\centering
\begin{tabular}{ccc}
\includegraphics[height=0.25\textwidth]{"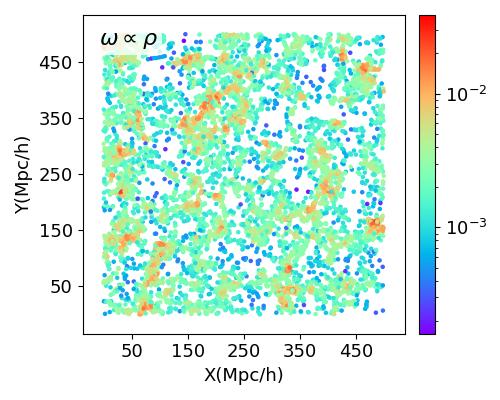"} &
\includegraphics[height=0.25\textwidth]{"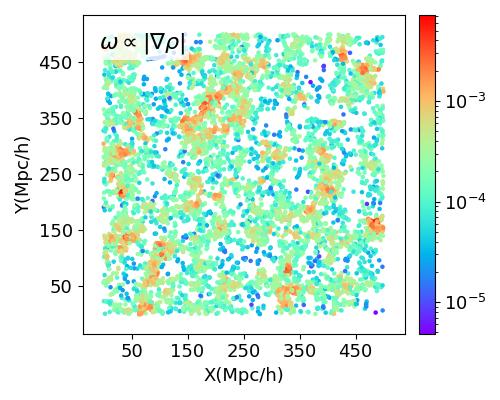"} &
\includegraphics[height=0.25\textwidth]{"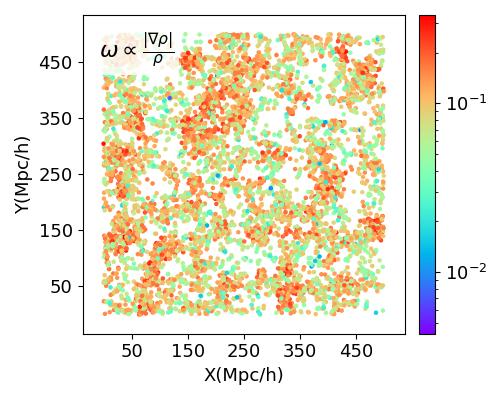"} \\
\end{tabular}
\caption{
Environmental quantities in a $500\times500~(h^{-1}{\rm Mpc})^2$ slice of a \textsc{Kun} halo catalog.
Left: the local-density mark $\rho_{n_{\rm NB}}$, which emphasizes compact overdensities.
Middle: the unnormalized gradient $|\nabla\rho_{n_{\rm NB}}|$, which remains visually correlated with the density field because the largest gradients occur around dense structures.
Right: the normalized gradient $|\nabla\rho_{n_{\rm NB}}|/\rho_{n_{\rm NB}}$, which reduces the direct density weighting and highlights relative environmental transitions.
}
\label{fig:slice}
\end{figure*}


Figures~\ref{fig:slice} illustrate the motivation for using the normalized gradient rather than the raw gradient. The raw gradient is still strongly tied to high-density regions, whereas $|\nabla\rho_{n_{\rm NB}}|/\rho_{n_{\rm NB}}$ suppresses much of this density scaling. The two adopted marks therefore probe complementary environmental information: one traces dense regions directly, and the other emphasizes relative environmental variation.

\subsubsection{Marked two-point statistic}

We now incorporate the environmental marks into a two-point statistic. For a given mark $w$ and mark power $\alpha$, we define the marked correlation function as
\begin{equation}
W^{\alpha}(\mathbf{r})
=
\left\langle
\delta(\mathbf{x}) w^{\alpha}(\mathbf{x})
\,
\delta(\mathbf{x}+\mathbf{r}) w^{\alpha}(\mathbf{x}+\mathbf{r})
\right\rangle .
\label{eq:MCF}
\end{equation}
Here $\delta$ is the tracer overdensity field, and the exponent $\alpha$ controls how strongly the mark affects the pair weighting.

For comparison, the ordinary two-point correlation function is
\begin{equation}
\xi(\mathbf{r})
=
\left\langle
\delta(\mathbf{x})
\delta(\mathbf{x}+\mathbf{r})
\right\rangle .
\label{eq:tpCF}
\end{equation}
Thus, $\alpha=0$ removes the mark weighting and recovers the standard unweighted two-point correlation function.

In redshift space, we estimate the marked statistic using the Landy-Szalay estimator,
\begin{equation}
W^{\alpha}(s,\mu)
=\frac{
WW(s,\mu)-2WR(s,\mu)+RR(s,\mu)
}{
RR(s,\mu)
}\,.
\label{eq:ls_mcf}
\end{equation}
Here $s$ is the pair separation and $\mu$ is the cosine of the angle between the pair separation vector and the line of sight. The quantities $WW$, $WR$, and $RR$ denote normalized weighted data-data, data-random, and random-random pair counts, respectively. Random points are assigned unit mark, following the convention adopted in previous MCF analyses~\citep{MCF_Yang,Xiao2022MCF}.

The sign and magnitude of $\alpha$ determine which environments dominate the weighted pair counts. Positive values, such as $\alpha=0.3$ and $\alpha=0.5$, upweight high-mark regions. Negative values, such as $\alpha=-0.3$ and $\alpha=-0.5$, increase the relative contribution of low-mark regions. Therefore, varying $\alpha$ allows the statistic to scan different environmental regimes.

\subsubsection{Scale and angular compression}

The full two-dimensional statistic $W^{\alpha}(s,\mu)$ contains both scale dependence and line-of-sight angular dependence. To build a compact data vector, we compress it in two complementary ways.

First, to isolate the scale dependence, we integrate over the angular range,
\begin{equation}
W^{\alpha}(s)
=
\int_{\mu_{\min}}^{\mu_{\max}}
W^{\alpha}(s,\mu)\,d\mu .
\label{eq:intMs}
\end{equation}
This statistic measures how the marked clustering amplitude varies with separation $s$ after averaging over the selected angular range.

Second, to isolate the angular dependence, we integrate over the separation range,
\begin{equation}
W^{\alpha}_{\Delta s}(\mu)
=
\int_{s_{\min}}^{s_{\max}}
W^{\alpha}(s,\mu)\,ds .
\label{eq:intMmu}
\end{equation}
This statistic measures how the marked signal changes with orientation relative to the line of sight. It is therefore sensitive to anisotropic redshift-space effects.

Unless stated otherwise, all analyses use $s_{\min}=15~h^{-1}{\rm Mpc}$, $s_{\max}=85~h^{-1}{\rm Mpc}$, and $0<\mu<0.8$.
With bin widths of $\Delta s = 10~h^{-1}{\rm Mpc}$ and $\Delta \mu = 0.2$. These choices give 7 separation bins and 4 angular bins for each mark and each value of $\alpha$.

\subsubsection{Normalized marked correlation functions}

For likelihood inference, we focus on the shape information rather than the overall integral of each marked statistic. We therefore define the normalized scale-dependent statistic as
\begin{equation}
\widehat{W}^{\alpha}(s)
=
\frac{
W^{\alpha}(s)
}{
\int_{s_{\min}}^{s_{\max}}
W^{\alpha}(s')\,ds'
}.
\label{eq:normMs}
\end{equation}
Similarly, the normalized angular statistic is
\begin{equation}
\widehat{W}^{\alpha}_{\Delta s}(\mu)
=
\frac{
W^{\alpha}_{\Delta s}(\mu)
}{
\int_{\mu_{\min}}^{\mu_{\max}}
W^{\alpha}_{\Delta s}(\mu')\,d\mu'
}.
\label{eq:normMmu}
\end{equation}

This normalization removes the total integrated amplitude of each statistic. The likelihood is therefore driven by changes in scale-dependent shape or angular-dependent shape. This choice is useful because it reduces sensitivity to an overall amplitude shift and emphasizes the relative redistribution of clustering signal across bins.

We build the final data vector by combining the normalized statistics from the two marks and five mark powers,
\begin{equation}
\alpha \in \{-0.5,-0.3,0,0.3,0.5\}.
\end{equation}
Here $\alpha$ controls the strength and direction of the environmental weighting. Positive values upweight high-mark regions, negative values give relatively more weight to low-mark regions, and $\alpha=0$ reduces to the ordinary unweighted two-point statistic.

\begin{figure*}[htpb]
\centering
\includegraphics[width=0.98\textwidth]{"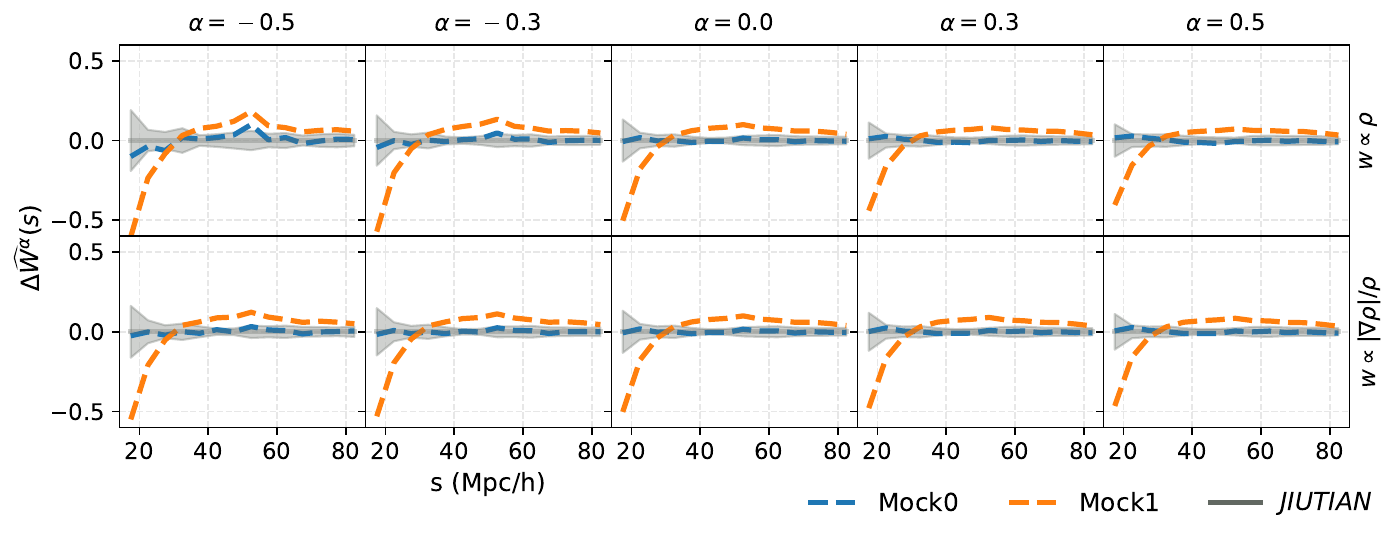"}
\caption{Scale-dependent marked statistic $\Delta \widehat{W}^{\alpha}(s)$ at $z=0.5$, defined as the difference between the \textsc{Kun} simulations and a $(600~h^{-1}{\rm Mpc})^3$ subsample of \textsc{Jiutian}. The gray shaded region shows the $1\sigma$ uncertainty. The blue and yellow dashed lines correspond to the fiducial (Mock0) and nonfiducial (Mock1) cosmologies from \textsc{Kun}, respectively. The upper and lower rows use $w=\rho_{n_{\rm NB}}$ and $w=|\nabla\rho_{n_{\rm NB}}|/\rho_{n_{\rm NB}}$, while columns show $\alpha=-0.5,-0.3,0,0.3,0.5$. The agreement between  \textsc{Jiutian} and the fiducial  \textsc{Kun} confirms consistency at fixed cosmology, whereas the nonfiducial case shows clear deviations. The distinct scale dependence across different $\alpha$ values motivates the joint multi-$\alpha$ analysis.}
\label{fig:xis_2pcf}
\end{figure*}

\begin{figure*}[htpb]
\centering
\includegraphics[width=0.98\textwidth]{"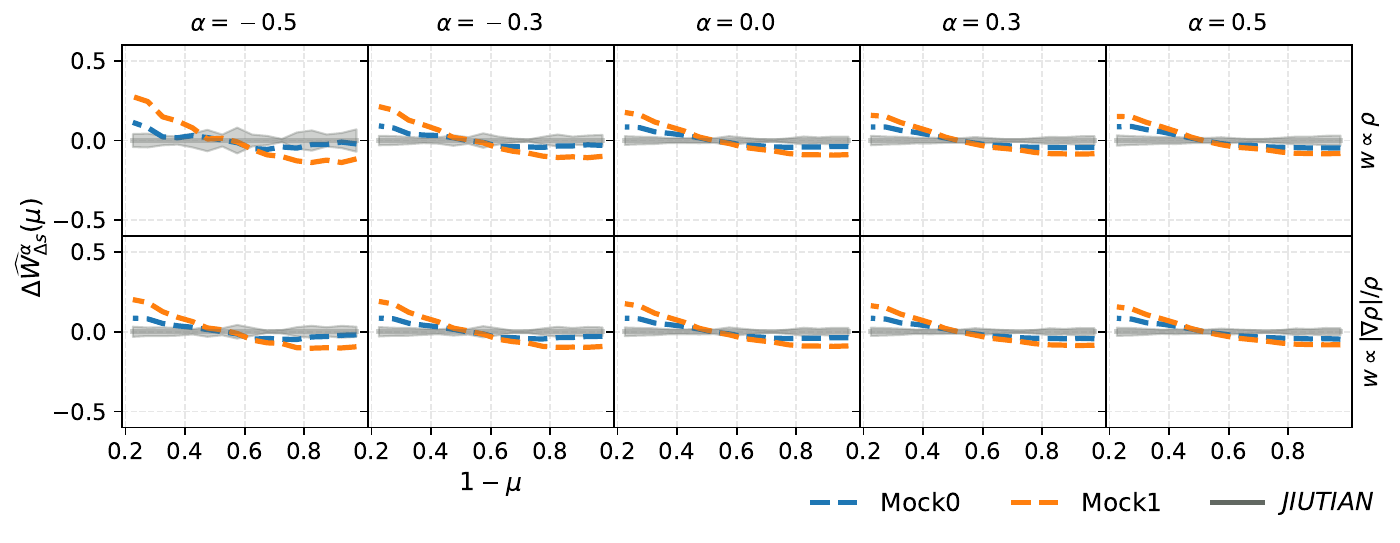"}
\caption{
Angular marked statistic $\Delta W^{\alpha}_{\Delta s}(\mu)$, shown as a function of $1-\mu$, for the same mark choices and simulations as in Figure~\ref{fig:xis_2pcf}. Compared with $W^{\alpha}(s)$, the dependence on $\alpha$ is less pronounced in amplitude, but the systematic variations across angular bins remain sensitive to cosmological parameters.}
\label{fig:ximu_2pcf}
\end{figure*}

Figures~\ref{fig:xis_2pcf} and~\ref{fig:ximu_2pcf} show the $\Delta \widehat{W}^{\alpha}(s)=\widehat{W}^{\alpha}_{kun}(s)-\widehat{W}^{\alpha}_{jiutian}(s)$ and $\Delta W^{\alpha}_{\Delta s}(\mu)=W^{\alpha}_{\Delta s,kun}(\mu)-W^{\alpha}_{\Delta s,jiutian}(\mu)$ statistics. The marked statistics do not only change in overall amplitude as $\alpha$ varies. Their scale dependence and angular dependence also change. This is the key reason for using a multi-mark and multi-$\alpha$ data vector: different environmental weightings respond differently to cosmological parameters and can help break degeneracies that remain in the ordinary 2PCF.

\subsection{Gaussian-process regression emulator}
\label{subsec:gpr}

We use Gaussian-process regression (GPR)~\citep{williams2006gaussian} to emulate the dependence of the marked correlation statistics on cosmological parameters. The goal is to obtain a fast and smooth prediction of the data vector at any cosmology within the training domain, without rerunning expensive simulations.

For a cosmological parameter vector $\hat{\theta}$, we denote one component of the MCF data vector by $w(\hat{\theta})$. In practice, the same procedure is applied independently to each component of $\widehat{W}^{\alpha}(s)$ and $\widehat{W}^{\alpha}_{\Delta s}(\mu)$. A Gaussian process places a prior distribution over functions,
\begin{equation}
f(\hat{\theta})
\sim
\mathcal{GP}
\left[
m(\hat{\theta}),
k(\hat{\theta},\hat{\theta}')
\right]\,,
\end{equation}
where $m(\hat{\theta})$ is the mean function and $k(\hat{\theta},\hat{\theta}')$ is the covariance kernel. The simulated statistic is modeled as
\begin{equation}
w(\hat{\theta}) = f(\hat{\theta}) + \epsilon,
\qquad
\epsilon \sim \mathcal{N}(0,\sigma_n^2),
\end{equation}
where $\sigma_n^2$ represents the emulator noise term.

Given training cosmologies $\hat{\theta}$ and a test cosmology $\hat{\theta}_*$, the joint distribution of the training values $f$ and the test value $f_*$ is
\begin{equation}
\begin{bmatrix}
f \\
f_*
\end{bmatrix}
\sim
\mathcal{N}
\left[
\begin{bmatrix}
m(\hat{\theta}) \\
m(\hat{\theta}_*)
\end{bmatrix},
\begin{bmatrix}
K(\hat{\theta},\hat{\theta})+\sigma_n^2 I
&
K(\hat{\theta},\hat{\theta}_*) \\
K(\hat{\theta}_*,\hat{\theta})
&
K(\hat{\theta}_*,\hat{\theta}_*)
\end{bmatrix}
\right]\,,
\end{equation}
where $K(\hat{\theta},\hat{\theta})$ is the kernel matrix evaluated over all training cosmologies.

Conditioning on the training simulations gives a Gaussian predictive distribution,
\begin{equation}
f_* \mid \mathcal{D}, \hat{\theta}_*
\sim
\mathcal{N}
\left(
\bar{f}_*,
{\rm Cov}(f_*)
\right)\,,
\end{equation}
where $\mathcal{D}=\{\hat{\theta},w\}$ denotes the training cosmologies and their simulated statistics. The predictive mean is
\begin{equation}
\bar{f}_*
=
m(\hat{\theta}_*)
+
K(\hat{\theta}_*,\hat{\theta})
\left[
K(\hat{\theta},\hat{\theta})+\sigma_n^2 I
\right]^{-1}
\left[
w-m(\hat{\theta})
\right]\,,
\end{equation}
and the predictive covariance is
\begin{equation}
{\rm Cov}(f_*)
=
K(\hat{\theta}_*,\hat{\theta}_*)
-
K(\hat{\theta}_*,\hat{\theta})
\left[
K(\hat{\theta},\hat{\theta})+\sigma_n^2 I
\right]^{-1}
K(\hat{\theta},\hat{\theta}_*)\,.
\end{equation}
The predictive mean $\bar{f}_*$ is used as the emulator prediction, while the predictive covariance quantifies the interpolation uncertainty.

We compare three kernel choices. The first is the radial basis function (RBF) kernel,
\begin{equation}
k_{\rm RBF}(\hat{\theta}_i,\hat{\theta}_j)
=
\sigma_f^2
\exp
\left(
-\frac{r_{ij}^2}{2l^2}
\right),
\end{equation}
where
\begin{equation}
r_{ij}
=
\left\|
\hat{\theta}_i-\hat{\theta}_j
\right\|\,.
\end{equation}
The RBF kernel assumes a very smooth dependence on cosmological parameters.

The second is the Matérn kernel with $\nu=5/2$ (M52),
\begin{equation}
k_{\rm M52}(\hat{\theta}_i,\hat{\theta}_j)
=
\sigma_f^2
\left(
1
+
\frac{\sqrt{5}r_{ij}}{l}
+
\frac{5r_{ij}^2}{3l^2}
\right)
\exp
\left(
-\frac{\sqrt{5}r_{ij}}{l}
\right)\,.
\end{equation}
Compared with the RBF kernel, the M52 kernel allows less restrictive smoothness while still producing sufficiently regular functions. This flexibility is useful when the statistic varies nontrivially across cosmological parameter space. Previous work has also found that the M52 kernel can outperform the RBF kernel for cosmological emulation tasks~\citep{Zhang2023M52}.

The third kernel is a product kernel (M52-RBF),
\begin{equation}
k_{\rm M52\text{-}RBF}(\hat{\theta}_i,\hat{\theta}_j)
=
k_{\rm M52}(\hat{\theta}_i,\hat{\theta}_j)
k_{\rm RBF}(\hat{\theta}_i,\hat{\theta}_j)\,.
\end{equation}
This composite kernel combines the flexibility of the M52 kernel with the smoothness of the RBF kernel, and is included to test whether a more structured covariance model improves emulator accuracy.

In the following analysis, we train the GPR emulator on the \textsc{Kun} simulation suite and validate its predictions using independent \textsc{Jiutian} simulations. The comparison among kernels allows us to identify a stable emulator model for the normalized MCF data vectors used in the likelihood analysis.

In practice, we include an overall amplitude parameter,
\begin{equation}
K(\hat{\theta}_i,\hat{\theta}_j)
=
A_{c}
\,k(\hat{\theta}_i,\hat{\theta}_j)\,,
\label{eq:gpr_full_kernel}
\end{equation}
where $A_c$ rescales the covariance amplitude, and $k$ denotes one of the RBF, M52, or M52-RBF kernels.

The training process in GPR is to find the optimal hyperparameters for the given training data and kernel function. This is achieved by the optimization of the log marginal likelihood of the input data via maximizing

\begin{equation}
\begin{aligned}
\ln \mathcal{L}_{\rm GPR} = & -\frac{1}{2} f^\top [K(\hat{\theta}, \hat{\theta}) + \sigma_n^2 I]^{-1} f \\
& - \frac{1}{2} \log |K(\hat{\theta}, \hat{\theta}) + \sigma_n^2 I| - \frac{n}{2} \log 2\pi\,.
\end{aligned}
\end{equation}
Here $n$ is the dimension of cosmological parameter space. We implement the GPR emulator with the Python package \textsc{scikit-learn}~\citep{scikit-learn}. The emulator is trained separately for each component of the normalized MCF data vector and then used to predict the statistics at new cosmological parameters.

We compare the RBF, M52, and M52-RBF kernels using the independent \textsc{Jiutian} validation simulations. All three kernels provide accurate predictions, and their validation errors differ by less than $5\%$. This indicates that the emulator performance is not sensitive to the kernel choice for the present data vector. We therefore adopt the RBF kernel as the fiducial emulator in the following analysis, because it is simple, stable, and sufficient for the required interpolation accuracy.

\begin{figure*}[htpb]
	\centering
	\begin{tabular}{cc}
		\includegraphics[width=0.48\textwidth]{"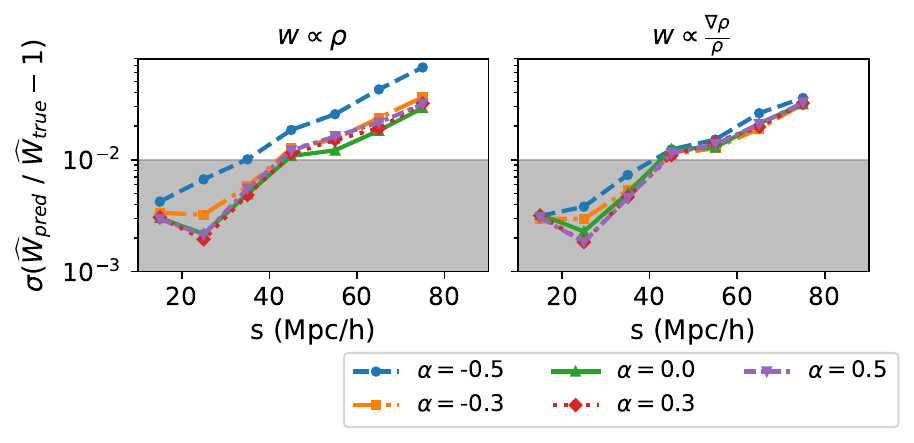"} &
		\includegraphics[width=0.48\textwidth]{"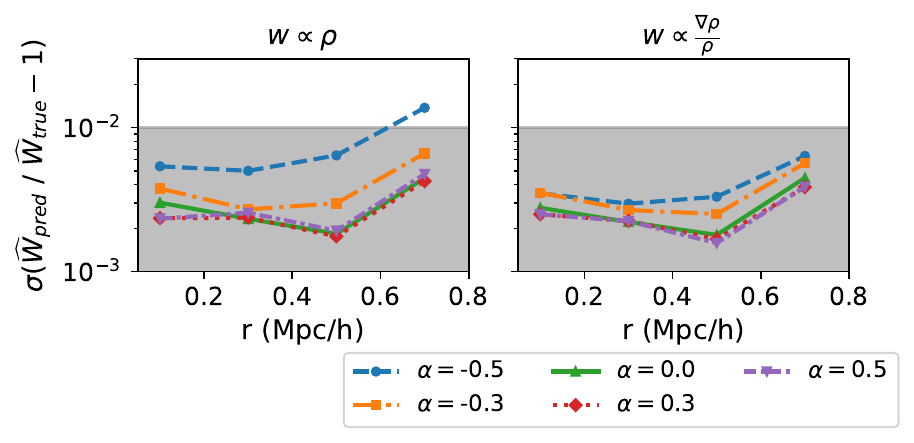"} \\
	\end{tabular}
	\caption{
	Leave-one-out validation errors of the GPR emulator.
	The left panel shows the 68th-percentile fractional error for the scale-dependent statistic $\widehat{W}^{\alpha}(s)$, while the right panel shows the corresponding error for the angular statistic $\widehat{W}^{\alpha}_{\Delta s}(\mu)$.
	The shaded regions indicate errors below $1\%$.
	Different curves correspond to different mark powers $\alpha$.
	}
	\label{fig:LOO}
\end{figure*}

To assess the generalization accuracy of the emulator, we perform leave-one-out (LOO) cross-validation. This test is useful for our relatively small simulation set because each simulation is used once as the validation sample, while all remaining simulations are used for training. Repeating this procedure over all simulations gives an estimate of the emulator error across the training cosmologies.

For each bin of the data vector, we define the LOO error as the 68th-percentile fractional error over all validation samples. As shown in Figure~\ref{fig:LOO}, the emulator achieves fractional errors below $8\%$ for the scale-dependent statistic $\widehat{W}^{\alpha}(s)$ and below $1.5\%$ for the angular statistic $\widehat{W}^{\alpha}_{\Delta s}(\mu)$ across the tested values of $\alpha$. These small errors indicate that the emulator accurately interpolates the MCF statistics across the cosmological parameter space covered by the training simulations.

We also observe that the errors for negative mark powers, especially $\alpha=-0.5$ and $\alpha=-0.3$, are generally larger than those for $\alpha=0$ and $\alpha>0$. This behavior is expected because negative $\alpha$ values give relatively more weight to low-mark environments. These regions contain fewer tracers and are therefore more affected by shot noise, leading to slightly larger emulator uncertainties.

\subsection{Covariance estimation and likelihood}
\label{subsec:covariance}
The Gaussian-process emulator provides predictions for the marked-correlation data vector at arbitrary cosmologies within the training range. We next describe the covariance matrix and likelihood used for parameter inference.
We denote the full data vector by $\mathbf{y}=\widehat{W}^{\alpha}(s)$ or $\mathbf{y}=\widehat{W}^{\alpha}_{\Delta s}(\mu)$, including all selected mark definitions, mark powers $\alpha$, and separation or angular bins. For a cosmological parameter vector $\boldsymbol{\theta}$, the emulator prediction is denoted by $\mathbf{y}_{\rm emu}(\boldsymbol{\theta})$. Assuming a Gaussian likelihood, we adopt
\begin{equation} 
\ln \mathcal{L}(\boldsymbol{\theta}) = -\frac{1}{2} \Delta\mathbf{y}^{\rm T} \mathbf{C}^{-1} \Delta\mathbf{y} + {\rm const.}\,, 
\label{eq:likelihood} 
\end{equation} where 
\begin{equation} 
\Delta\mathbf{y} = \mathbf{y}_{\rm obs} - \mathbf{y}_{\rm emu}(\boldsymbol{\theta})\,.
\end{equation}
Here $\mathbf{y}_{\rm obs}$ is the measured data vector, and $\mathbf{y}_{\rm emu}(\boldsymbol{\theta})$ is the phase-corrected emulator prediction defined below. Since the covariance matrix is fixed in our analysis, the normalization term of the Gaussian likelihood does not affect parameter inference and is absorbed into the constant.

The total covariance is modeled as
\begin{equation}
\mathbf{C}
=
\mathbf{C}_{\rm data}
+
\mathbf{C}_{\rm emu}\,.
\label{eq:total_cov}
\end{equation}
The first term describes the statistical uncertainty of the measured data vector, while the second term accounts for interpolation errors from the emulator. The effect of the fixed initial phase in the training simulations is treated as a correction to the emulator mean prediction, rather than as an additional covariance term.

\subsubsection{Data covariance}

We estimate the data covariance from subvolumes of the \textsc{Jiutian} simulation. The full simulation is divided into $N_{\rm sub}=5^3=125$ subsamples, each with volume
 $V_{\rm sub} = (400~h^{-1}{\rm Mpc})^3$. Let $\mathbf{y}_k$ be the data vector measured from the $k$-th subvolume, and let
\begin{equation}
\bar{\mathbf{y}}
=
\frac{1}{N_{\rm sub}}
\sum_{k=1}^{N_{\rm sub}}
\mathbf{y}_k
\end{equation}
be the mean over all subvolumes. The covariance for an observed volume $V_{\rm obs}$ is then estimated as
\begin{equation}
\mathbf{C}_{\rm data}
=
\frac{V_{\rm sub}}{V_{\rm obs}}
\frac{1}{N_{\rm sub}-1}
\sum_{k=1}^{N_{\rm sub}}
\left(
\mathbf{y}_k-\bar{\mathbf{y}}
\right)
\left(
\mathbf{y}_k-\bar{\mathbf{y}}
\right)^{\rm T}.
\label{eq:data_cov}
\end{equation}
The factor $V_{\rm sub}/V_{\rm obs}$ rescales the covariance from the subvolume size to the target observational volume.

\subsubsection{Emulator covariance}

The emulator covariance quantifies the uncertainty introduced by interpolation across cosmological parameter space. We estimate this term using the leave-one-out validation described in the previous section. For the $j$-th training cosmology, the emulator is trained on all other cosmologies and then evaluated at the omitted point. The residual vector is
\begin{equation}
\mathbf{e}_j
=
\mathbf{y}^{(-j)}_{\rm emu}(\boldsymbol{\theta}_j)
-
\mathbf{y}_{\rm true}(\boldsymbol{\theta}_j),
\end{equation}
where $\mathbf{y}^{(-j)}_{\rm emu}$ is the leave-one-out emulator prediction and $\mathbf{y}_{\rm true}$ is the statistic measured directly from the simulation.

The emulator covariance is estimated from these residuals:
\begin{equation}
\mathbf{C}_{\rm emu}
=
\frac{1}{N_{\rm train}-1}
\sum_{j=1}^{N_{\rm train}}
\left(
\mathbf{e}_j-\bar{\mathbf{e}}
\right)
\left(
\mathbf{e}_j-\bar{\mathbf{e}}
\right)^{\rm T},
\label{eq:emu_cov}
\end{equation}
where
\begin{equation}
\bar{\mathbf{e}}
=
\frac{1}{N_{\rm train}}
\sum_{j=1}^{N_{\rm train}}
\mathbf{e}_j .
\end{equation}
In practice, we find that $\mathbf{C}_{\rm emu}$ is much smaller than $\mathbf{C}_{\rm data}$, indicating that emulator interpolation errors are subdominant compared with the statistical uncertainty of the data vector.

\subsubsection{Correction for fixed initial phases}

The \textsc{Kun} training simulations are generated with fixed initial phases. While this suppresses sample variance and improves emulator training, it can introduce a systematic offset between the emulator prediction and the ensemble-averaged statistic from independent realizations. Following the ratio-based correction~\cite{Yuan_2022}, we account for this effect by rescaling the emulator mean prediction.

We define the bin-wise correction factor as
\begin{equation}
R_i=
\frac{\bar{y}_{{\rm Jiutian},i}}
{y_{{\rm Kun,fid},i}}\,,
\end{equation}
where $\bar{y}_{{\rm Jiutian},i}$ is the ensemble mean measured from the independent \textsc{Jiutian} realizations, and $y_{{\rm Kun,fid},i}$ is the fiducial measurement from \textsc{Kun}. We then rescale the emulator prediction as
\begin{equation}
y_{{\rm emu},i}(\boldsymbol{\theta}) \rightarrow
R_i\, y_{{\rm emu},i}(\boldsymbol{\theta})\,.
\label{eq:phase_correction}
\end{equation}
This procedure assumes that the fixed-phase bias is only weakly dependent on cosmology, so that the ratio $R_i$ estimated at the fiducial cosmology can be applied across parameter space. We therefore use $y_{\rm model}$ as the fiducial prediction in the likelihood analysis. Since the residual phase uncertainty is subdominant compared with the statistical covariance, no additional phase-covariance term is included.

The likelihood uses the full covariance matrix of the data vector, which fully captures correlations between all elements. Because the data vector combines multiple bins, marks, and mark powers, the covariance between elements is non-negligible and must be fully retained. Correlations arise both from neighboring separation (or angular) bins and from the two marks, which trace related properties of the same underlying density field.

\begin{figure*}[htpb]
\centering
\includegraphics[height=0.4\textwidth]{"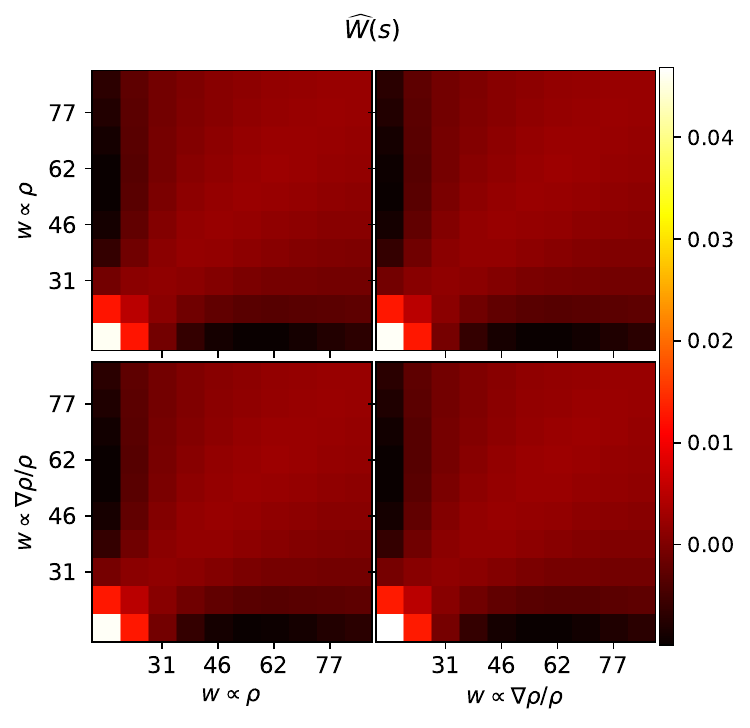"}
\includegraphics[height=0.4\textwidth]{"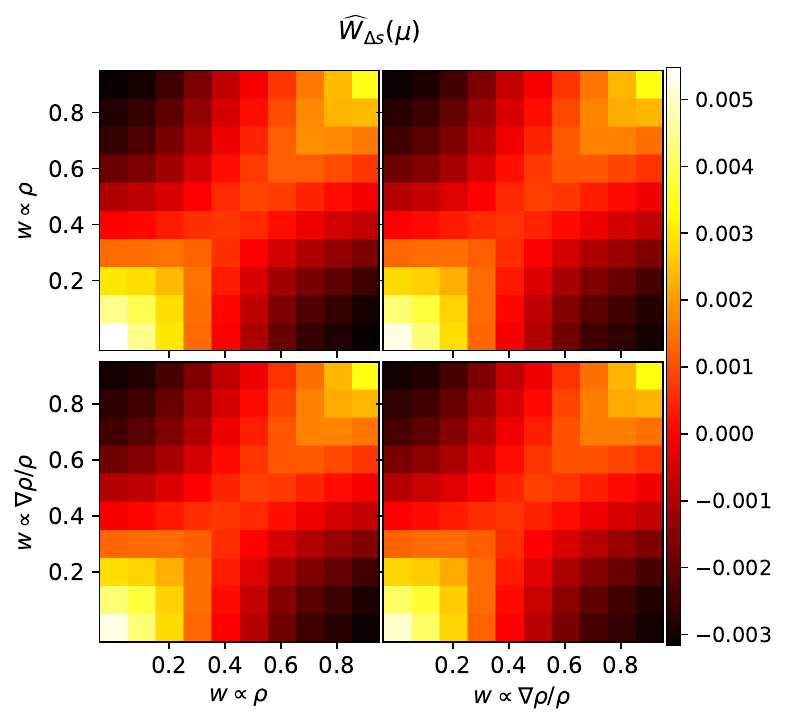"}
\caption{
Covariance matrices for $\widehat{W}^{\alpha}(s)$ (left) and $\widehat{W}^{\alpha}_{\Delta s}(\mu)$ (right) for the representative mark power $\alpha=0.5$. Strong within-mark correlations are visible, together with non-negligible cross-correlations between $w\propto\rho_{n_{\rm NB}}$ and $w\propto|\nabla\rho_{n_{\rm NB}}|/\rho_{n_{\rm NB}}$. The overall covariance amplitude is larger for $\widehat{W}^{\alpha}(s)$, with maxima of approximately $8\times10^{-3}$ and $1.25\times10^{-3}$, respectively.
}
\label{fig:cov}
\end{figure*}

Figure~\ref{fig:cov} shows substantial off-diagonal structure, indicating that many bins are strongly correlated. Increasing the number of bins would therefore not proportionally increase the independent information content, but would instead amplify noise in the estimated covariance and degrade matrix inversion. For this reason, we adopt a coarse binning scheme for both $\widehat{W}^{\alpha}(s)$ and $\widehat{W}^{\alpha}_{\Delta s}(\mu)$. This balances information retention against covariance stability, ensuring robust parameter inference with the available simulation volume.

\subsubsection{Inverse covariance correction}

Because the covariance matrix is estimated from a finite number of subvolumes, its inverse is biased. We correct this bias using the Hartlap factor~\citep{Hartlap_2006}. The debiased inverse covariance is
\begin{equation}
\widehat{\mathbf{C}}^{-1}
=
\frac{
N_{\rm sub}-N_d-2
}{
N_{\rm sub}-1
}
\mathbf{C}^{-1},
\label{eq:hartlap}
\end{equation}
where $N_d$ is the dimension of the data vector. This correction is well defined only when $N_{\rm sub}>N_d+2$, which motivates the compact binning scheme adopted for the MCF data vector.

\section{Results}
\label{sec:results}
\subsection{Comparison of 2PCF and MCFs}

\begin{figure*}[htpb]
	\centering
	\begin{tabular}{cc}
		\includegraphics[width=0.48\textwidth]{"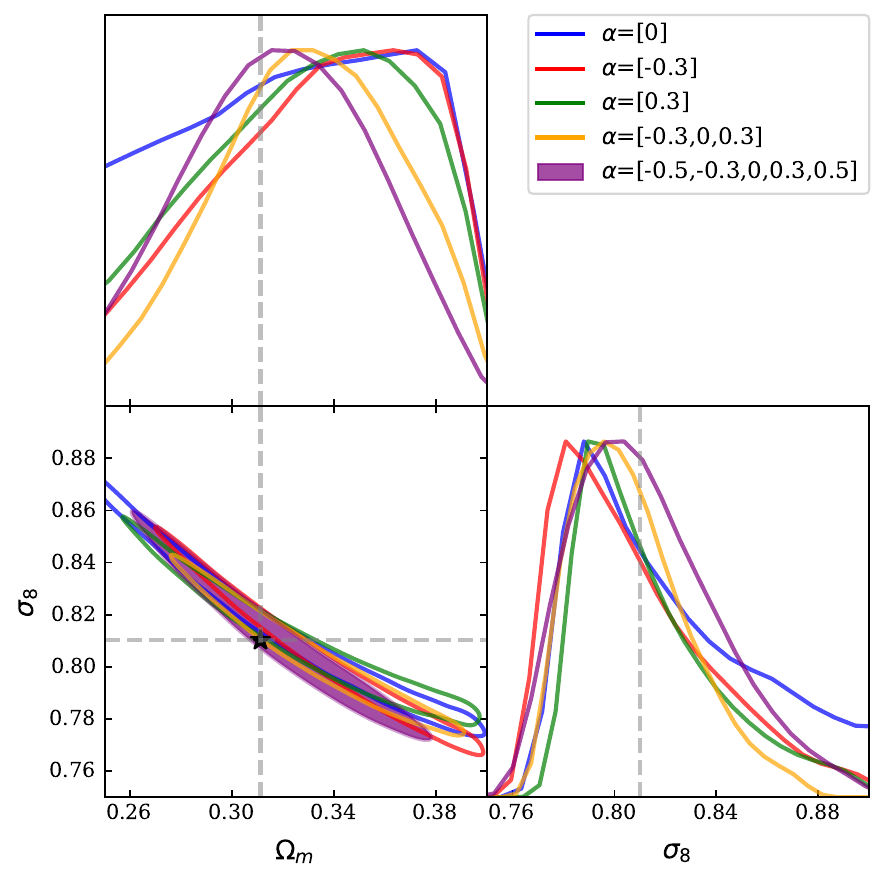"} & \includegraphics[width=0.48\textwidth]{"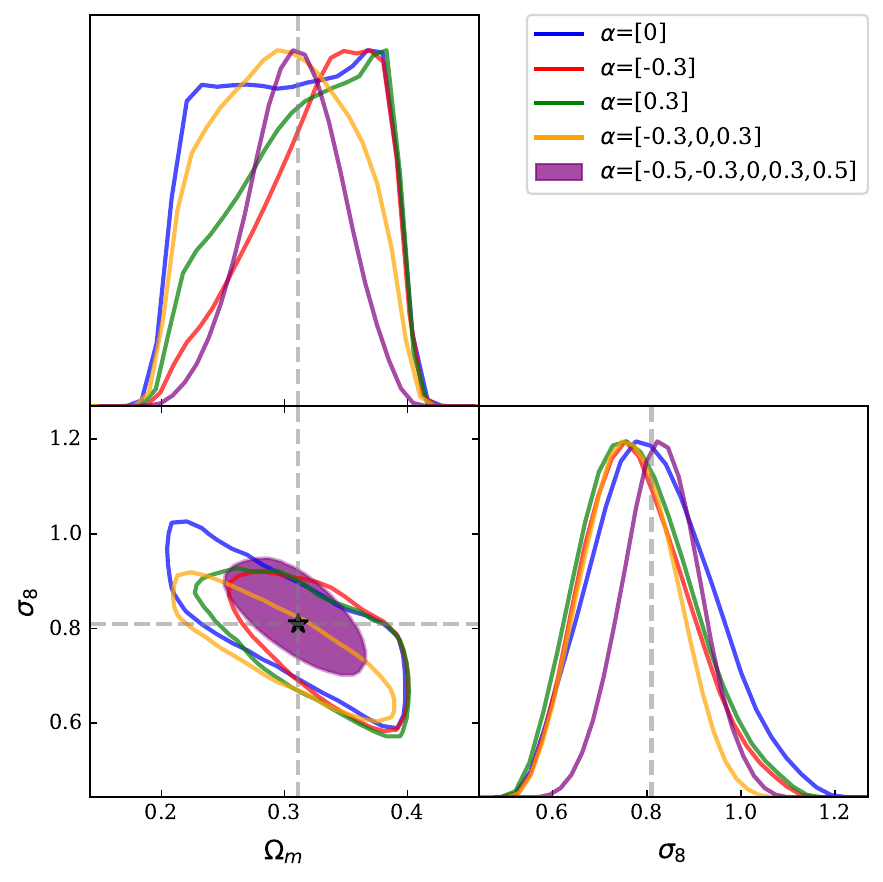"} \\
	\end{tabular}
\caption{Comparison of $1\sigma$ posterior contours for $\Omega_m$ and $\sigma_8$, derived from the standard 2PCF and MCFs with different $\alpha$ values.Dashed lines indicate the fiducial values $\Omega_m = 0.311$ and $\sigma_8 = 0.81$.}
\label{fig:mcmc_results_alphas}
\end{figure*}

\begin{figure}[htpb]
	\centering
		\includegraphics[width=0.48\textwidth]{"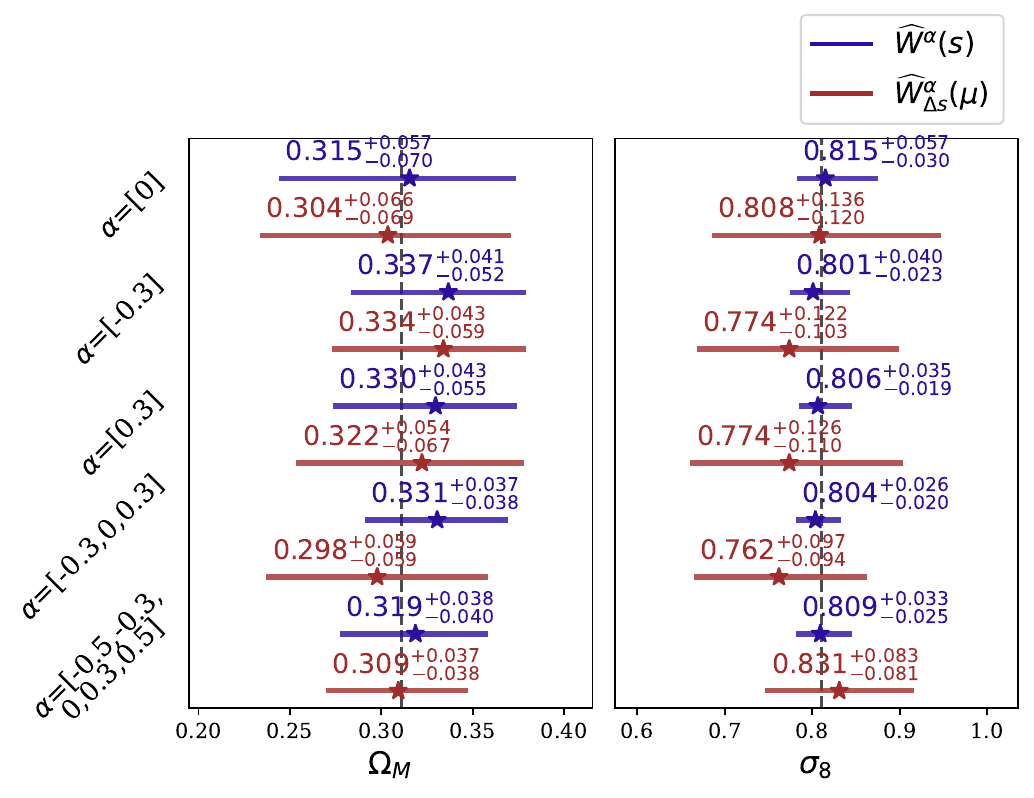"} 
\caption{1D posteriors for $\Omega_m$ and $\sigma_8$ from the 2PCF and MCFs (varying $\alpha$). Dashed lines mark the fiducial values.}
\label{fig:mcmc_results_alphas_1d}
\end{figure}

We quantify the constraining power of MCFs relative to the standard 2PCF, considering both individual marks and their combinations. Since correlation statistics are primarily sensitive to the matter density $\Omega_{m}$ and the fluctuation amplitude $\sigma_8$, and these two parameters exhibit a strong degeneracy in such measurements, we perform the cosmological inference in the $\Omega_m$--$\sigma_8$ plane.

As a validation dataset, we use a $(600\,h^{-1}{\rm Mpc})^3$ sub-volume of the \textsc{Jiutian} simulation, which provides an independent realization at the fiducial cosmology. Likelihood evaluations are performed using the emulator described in Section~\ref{sec:methodology}, and posterior sampling is carried out with the \textsc{emcee} package~\citep{emcee}.

Figure~\ref{fig:mcmc_results_alphas} presents the resulting $1\sigma$ constraints on $\Omega_m$ and $\sigma_8$, derived from the 2PCF and from MCFs with different choices of the mark exponent $\alpha$ (visualized using \textsc{getdist}~\citep{Lewis_2025}). Results are shown for both the scale-dependent statistic $\widehat{W}^{\alpha}(s)$ (left panel) and the angular statistic $\widehat{W}^{\alpha}_{\Delta s}(\mu)$ (right panel).

Several trends are apparent. First, a single marked statistic (fixed $\alpha$) provides constraints that are comparable to, but not significantly stronger than, the standard 2PCF ($\alpha=0$). Second, combining multiple $\alpha$ values leads to a clear improvement. In particular, both the three-weight combination $\alpha=\{-0.3,0,0.3\}$ and the full set $\alpha=\{-0.5,-0.3,0,0.3,0.5\}$ tighten the contours relative to the 2PCF alone, with the full combination yielding the strongest constraints.

This behavior reflects the complementary information encoded by different marks: positive $\alpha$ emphasizes overdense environments, while negative $\alpha$ enhances the contribution from underdense regions. Combining them probes clustering across a wide range of environments, helping to break degeneracies in the unweighted statistic. Overall, these results demonstrate that the gain from MCFs does not arise from any single weighting scheme, but from their joint use as a multi-component summary statistic.


The improvement is first evident in the one-dimensional posteriors shown in Figure~\ref{fig:mcmc_results_alphas_1d}. For $\widehat{W}^{\alpha}(s)$, combining three weights ($\alpha=\{-0.3,0,0.3\}$) reduces the $1\sigma$ uncertainties by $\sim41\%$ for $\Omega_m$ and $\sim47\%$ for $\sigma_8$ relative to the standard 2PCF. Including all five weights further changes the constraints, giving reductions of $\sim38\%$ and $\sim33\%$, respectively. Although the full five-weight combination yields slightly weaker statistical constraints than the three-weight case, it produces smaller shifts in the best-fit parameters relative to the true cosmology: the bias in $\Omega_m$ decreases from 0.20 to 0.08, and that in $\sigma_8$ from 0.007 to 0.002. For the angular statistic $\widehat{W}^{\alpha}_{\Delta s}(\mu)$, the corresponding reductions are $13\%$ and $25\%$ (three weights), and $45\%$ and $36\%$ (five weights). These results indicate that combining multiple $\alpha$ values substantially improves constraining power compared to any single statistic. Overall, we adopt the full five-weight combination as the baseline choice, as it provides a good balance between statistical precision and robustness.

To quantify joint constraints including parameter degeneracies, we adopt the Figure of Merit (FoM) defined as the inverse square root of the covariance determinant~\citep{Wang2008FOM},
\begin{equation}
\mathrm{FoM} \equiv \frac{1}{\sqrt{\det \mathbf{C}}}.
\end{equation}
For the two parameters ($\Omega_m$-$\sigma_8$), this becomes
\begin{equation}
\mathrm{FoM} = \frac{1}{\sigma(\Omega_m)\,\sigma(\sigma_8)\,\sqrt{1-\rho^2}},
\end{equation}
where $\rho$ denotes the correlation coefficient between $\Omega_m$ and $\sigma_8$. This form directly reflects the inverse area of the error ellipse and generalizes to higher dimensions via $\det\mathbf{C}$.

Table~\ref{tab:weight_alphas} summarizes the FoM for different $\alpha$ configurations, normalized to the five-weight $\widehat{W}^{\alpha}(s)$ case.  Compared to the 2PCF-only case ($\alpha=0$), the three-weight combination ($\alpha=[-0.3,0,0.3]$) increases the FoM by factors of $\sim2.51$ for $\widehat{W}^{\alpha}(s)$ and $\sim1.70$ for $\widehat{W}^{\alpha}_{\Delta s}(\mu)$. Including all five weights ($\alpha=[-0.5,-0.3,0,0.3,0.5]$) yields FoM gains of $\sim1.90$ and $\sim2.39$, respectively. We find that for $\widehat{W}^{\alpha}(s)$, the configuration using a single negative weight ($\alpha=-0.3$) yields a larger FoM than the full five-weight combination. However, as shown in Figure~\ref{fig:mcmc_results_alphas}, this case exhibits stronger parameter degeneracies, resulting in biased posteriors where the $1\sigma$ contours do not fully recover the true cosmology.

Taken together, these results show that the primary gain from MCFs arises from combining multiple environmental weightings, which effectively reduces both parameter uncertainties and degeneracies, leading to a substantially smaller allowed parameter volume.

\begin{table}[htbp]
\centering
\label{tab:weight_alphas}
\begin{tabular}{@{\extracolsep{\fill}} lcc}
\hline
 & $\widehat{W}^{\alpha}_{\Delta s}(\mu)$ & $\widehat{W}^{\alpha}(s)$ \\[1.5ex]
\hline
$\alpha=[0]$               & 0.044 & 0.525 \\
$\alpha=[-0.3]$            & 0.059 & 1.060  \\
$\alpha=[0.3]$             & 0.049 & 0.788 \\
$\alpha=[-0.3,0,0.3]$      & 0.075 & 1.320  \\
$\alpha=[-0.5,-0.3,0,0.3,0.5]$ & 0.105  & 1.000  \\[1.5ex]
\hline
\end{tabular}
\caption{FoM for various $\alpha$ values, normalized to the $\widehat{W}^{\alpha}(s)$ FoM for the fiducial combination $\alpha=\{-0.5,-0.3,0,0.3,0.5\}$.}
\end{table}

The composite parameter
\begin{equation}
S_8 \equiv \sigma_8 \left(\frac{\Omega_m}{0.3}\right)^{1/2}\,,
\end{equation}
shows a similar trend. For the standard 2PCF ($\alpha=0$), we obtain
$S_8 = 0.804\pm 0.093$, consistent with the fiducial value
$S_8^{\mathrm{fid}}\simeq 0.825$ within $1\sigma$. Single marked statistics with $\alpha=-0.3$ and $\alpha=0.3$ yield comparable central values and uncertainties. However, the three-weight combination $\alpha=\{-0.3,0,0.3\}$ shifts the posterior to
$S_8 = 0.75\pm{0.061}$, corresponding to a $\sim2\sigma$ deviation from the fiducial value, despite the reduced statistical uncertainty. In contrast, the full five-weight combination
$\alpha=\{-0.5,-0.3,0,0.3,0.5\}$
recovers
$S_8 = 0.836\pm0.062$, in excellent agreement with the fiducial cosmology, while still achieving a $\sim33\%$ reduction in the $1\sigma$ uncertainty relative to the 2PCF.

These results indicate that although combining a small number of marked statistics improves precision, it can introduce systematic shifts in degenerate parameter combinations such as $S_8$. A wider range of $\alpha$ values is therefore required to achieve both high precision and unbiased parameter recovery.

\subsection{Impact of weighting scheme and scale cuts on MCF constraints}
In this section, we study how the cosmological constraints from MCFs depend on key analysis choices, in particular the weighting scheme and the range of separation scales $s$. Unless otherwise specified, all results use the five-weight combination $\alpha = [-0.5, -0.3, 0, 0.3, 0.5]$.


\begin{figure*}[htpb]
	\centering
	\begin{tabular}{cc}
		\includegraphics[width=0.48\textwidth]{"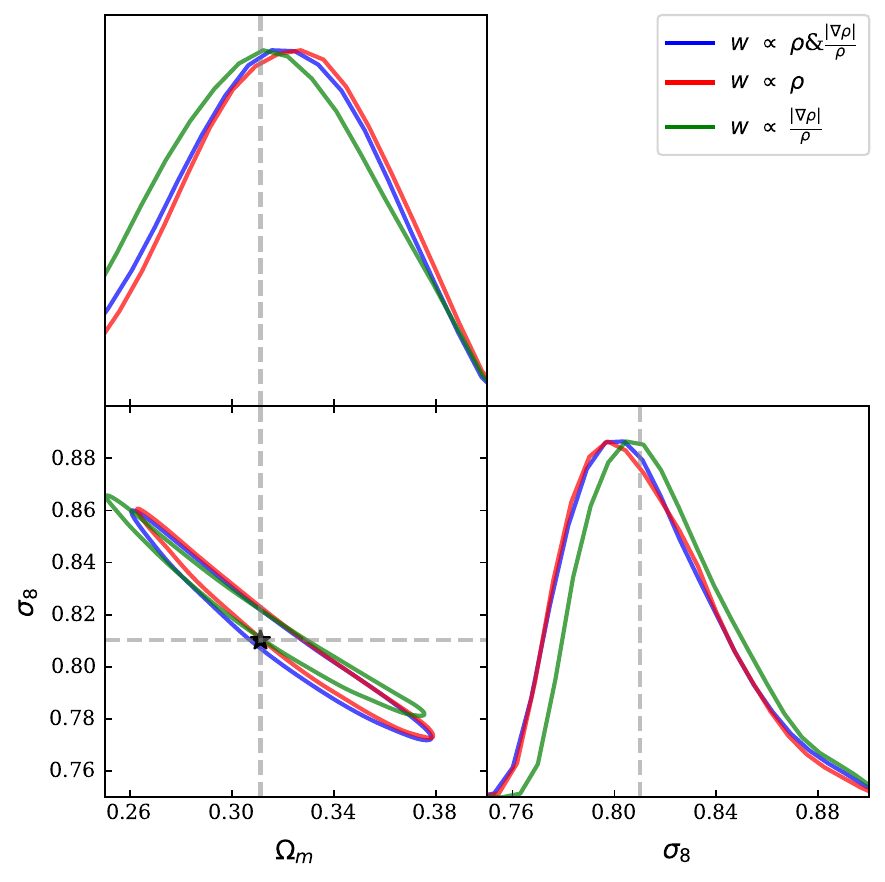"} & \includegraphics[width=0.48\textwidth]{"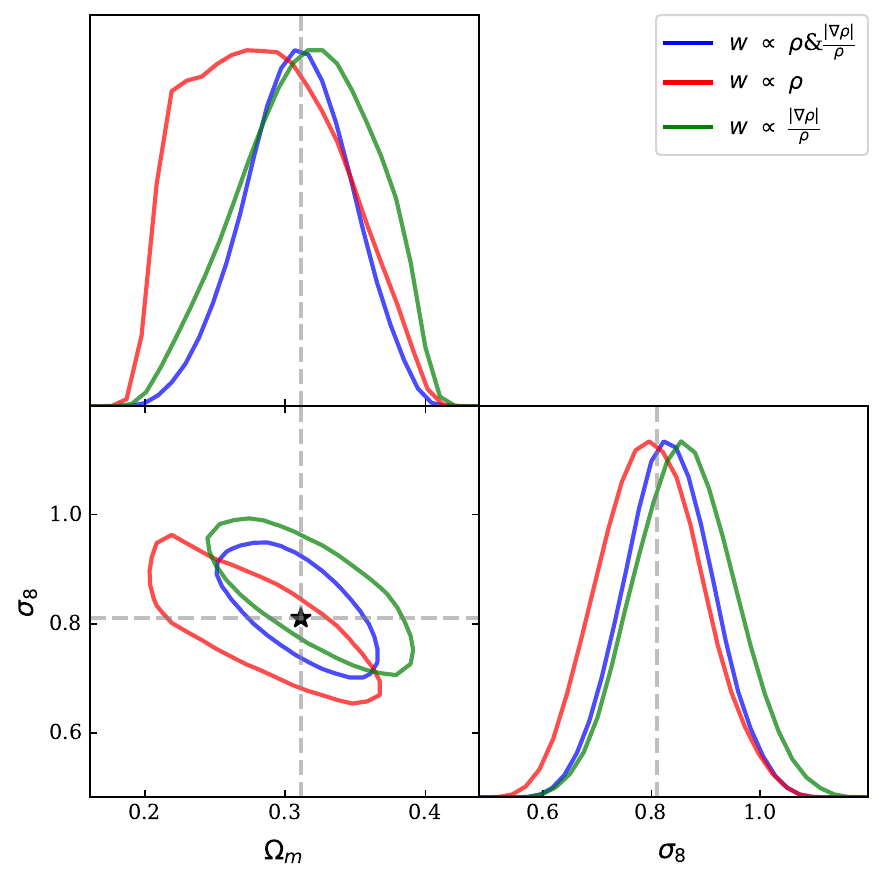"} \\
	\end{tabular}
\caption{The MCMC results for $\widehat{W}^{\alpha}(s)$ (left) and $\widehat{W}^{\alpha}_{\Delta s}(\mu)$ (right). For all cases, we adopt the same set of mark parameters $\alpha = [-0.5, -0.3, 0, 0.3, 0.5]$, and perform a joint combination of the corresponding contributions across these $\alpha$ values. We compare different weighting schemes, including $w \propto \rho$ and $w \propto \nabla\rho/\rho$, all evaluated under the fiducial binning scheme.} 
\label{fig:mcmc_results_weight_keys}
\end{figure*}

\begin{figure*}[htpb]
	\centering
	\begin{tabular}{cc}
		\includegraphics[width=0.48\textwidth]{"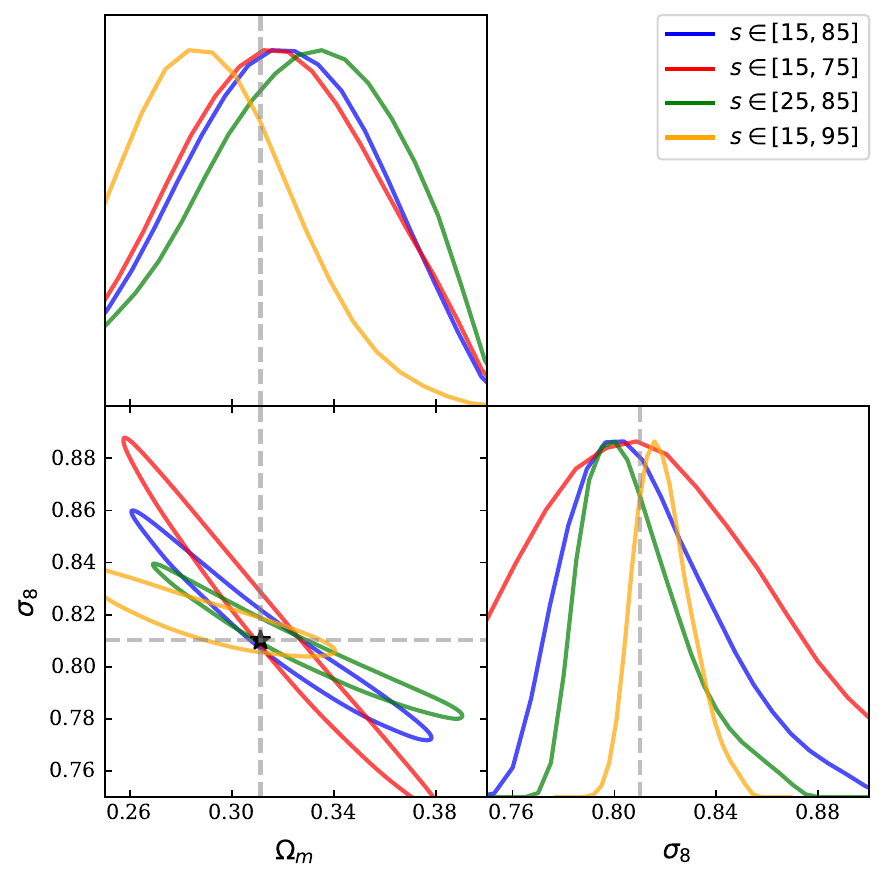"} & \includegraphics[width=0.48\textwidth]{"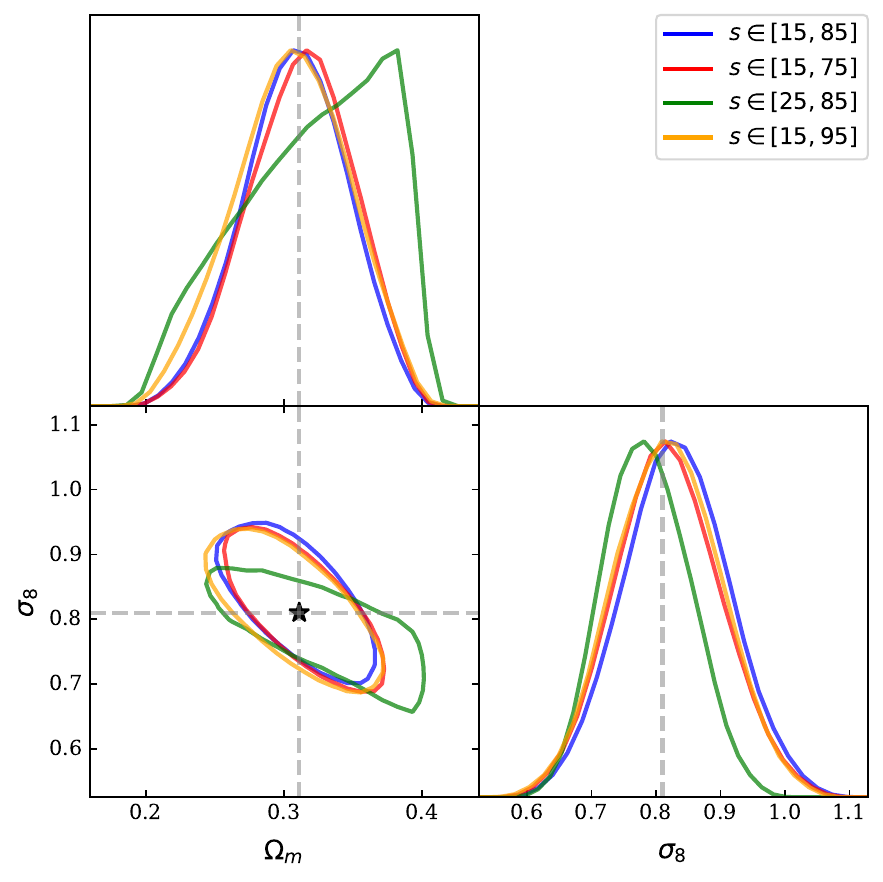"} \\
	\end{tabular}
\caption{Similar to Figure~\ref{fig:mcmc_results_weight_keys}, but showing results for different choices of the separation range $s$: $[15, 75]$, $[15, 85]$, $[25, 85]$, and $[15, 95]~h^{-1}\mathrm{Mpc}$. The binning scheme is kept fixed with the fiducial choices $\Delta s = 10~h^{-1}\mathrm{Mpc}$ and $\mu \in [0, 0.8]$ with $\Delta \mu = 0.2$, respectively.}
\label{fig:mcmc_results_s_range}
\end{figure*}

\begin{figure}[htpb]
	\centering
	\includegraphics[width=0.48\textwidth]{"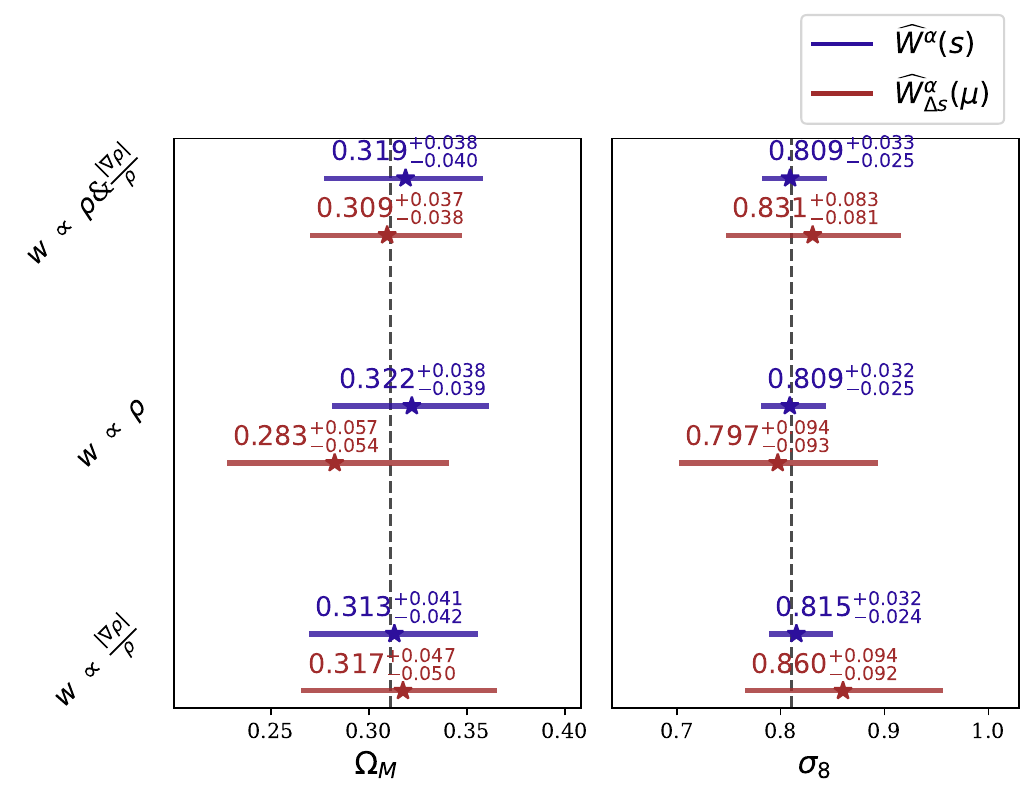"}
\caption{Same configurations as in Figure~\ref{fig:mcmc_results_weight_keys}, but presenting the marginalized one-dimensional MCMC constraints.}
\label{fig:mcmc_results_weight_keys_1d}
\end{figure}

\begin{figure}[htpb]
	\centering
	\includegraphics[width=0.48\textwidth]{"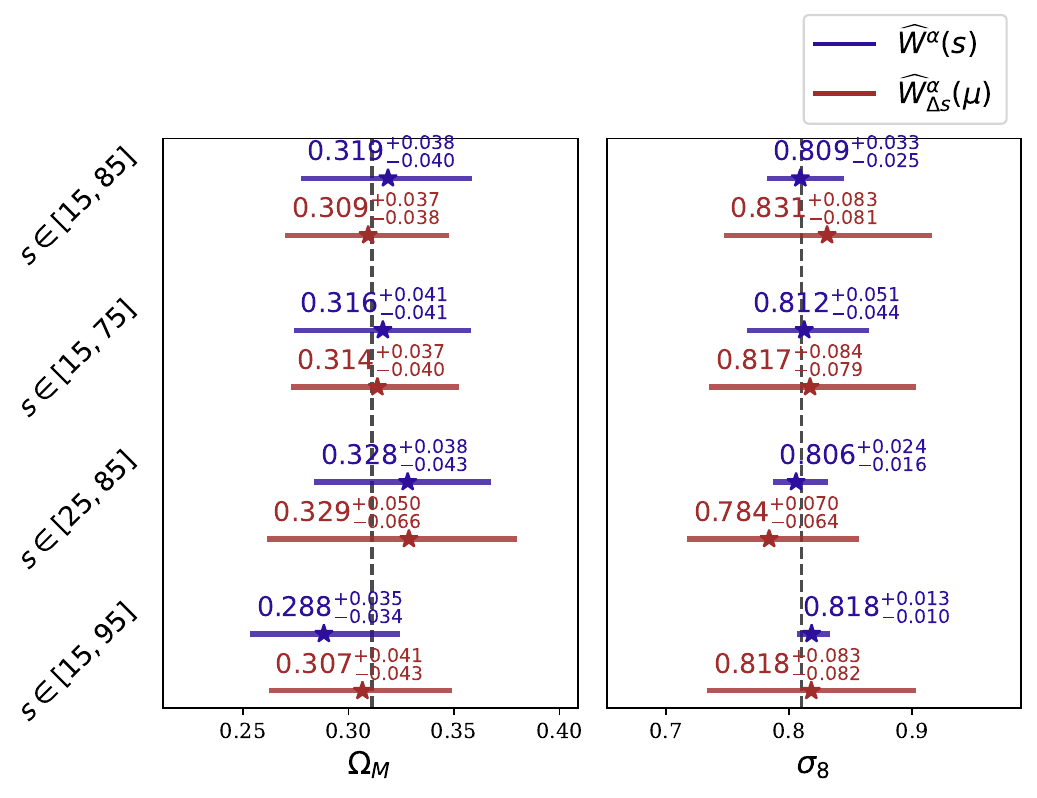"}
\caption{Same configurations as in Figure~\ref{fig:mcmc_results_s_range}, but presenting the one-dimensional marginalized MCMC constraints.}
\label{fig:mcmc_results_s_range_1d}
\end{figure}

\begin{table}[htbp]
\centering
\caption{FoM for different marker schemes. All FoM values are normalized to that of $\widehat{W}^{\alpha}(s)$ with the joint weighting configuration $w \propto |\nabla \rho|/\rho$ and $w \propto \rho$, which is set to unity for reference.}
\label{tab:weight_keys}
\begin{tabular}{lccc}
\hline
 & $w \propto |\nabla \rho|/\rho$ & $w \propto \rho$ & Joint\\
\hline
$\widehat{W}^{\alpha}_{\Delta s}(\mu)$ & 0.078 & 0.073 & 0.105 \\
$\widehat{W}^{\alpha}(s)$               & 1.086 & 1.168 & 1.000 \\
\hline
\end{tabular}
\end{table}

\begin{table}[htbp]
\centering
\caption{FoM for different choices of the separation range $s$. All values are normalized to the FoM of the isotropic statistic $\widehat{W}^{\alpha}(s)$ with $s \in [15,65]~h^{-1}{\rm Mpc}$, which is set to unity.}
\label{tab:s_range}
\begin{tabular}{lcccc}
\hline
$s\in $ & $[15,85]$ & $[15,75]$ & $[25,85]$ & $[15,95]$ \\
\hline
$\widehat{W}^{\alpha}_{\Delta s}(\mu)$ & 0.105 & 0.110 & 0.094 & 0.098 \\
$\widehat{W}^{\alpha}(s)$               & 1.00  & 0.755 & 1.39  & 1.30  \\
\hline
\end{tabular}
\end{table}

Figure~\ref{fig:mcmc_results_weight_keys} compares the cosmological constraints obtained using two weighting schemes, $w \propto \rho$ and $w \propto |\nabla\rho|/\rho$, as well as their joint combination.  For the isotropic statistic $\widehat{W}^{\alpha}(s)$, the joint analysis yields constraints on $\Omega_m$ and $\sigma_8$ that are comparable to those obtained from the best single-weight case, as quantified by the FoM values in Table~\ref{tab:weight_keys}. The joint FoM is approximately $14\%$ lower than that of $w \propto \rho$ alone, indicating that the two weighting schemes are strongly correlated in the isotropic MCF and therefore provide largely redundant information. Nevertheless, the joint combination slightly improves the constraint on $\sigma_8$, while leading to a marginally larger shift in $\Omega_m$ toward the fiducial value, reflecting a partial reduction in parameter degeneracies.

A much stronger improvement is found when using the anisotropic statistic $\widehat{W}^{\alpha}_{\Delta s}(\mu)$. In this case, the joint analysis tightens the marginalized constraints by approximately $35\%$ for $\Omega_m$ and $11\%$ for $\sigma_8$, as shown in Figure~\ref{fig:mcmc_results_weight_keys_1d}. Consistently, the FoM increases by a factor of $\sim1.4$ compared with either individual weighting scheme. These results indicate that the anisotropic MCF captures more complementary information from different environmental weights. In other words, once the angular dependence with respect to the line of sight is retained, the two weighting schemes become less redundant and their combination substantially improves the overall constraining power.

We next examine how the constraints depend on the separation range $s$, as shown in Figures~\ref{fig:mcmc_results_s_range} and~\ref{fig:mcmc_results_s_range_1d}. We consider four configurations: $s\in [15,75]$, $[15,85]$, $[15,95]$, and $[25,85]~h^{-1}{\rm Mpc}$.

For the isotropic statistic $\widehat{W}^{\alpha}(s)$, the constraining power depends sensitively on the adopted separation range. Reducing $s_{\max}$ from $85$ to $75~h^{-1}{\rm Mpc}$ leads to a significant decrease in the FoM of approximately $24\%$ (see Table~\ref{tab:s_range}). In contrast, either increasing $s_{\min}$ from $15$ to $25~h^{-1}{\rm Mpc}$  or extending $s_{\max}$ to $95~h^{-1}{\rm Mpc}$  improves the FoM by $\sim39\%$ and $\sim30\%$, respectively. However, these gains in statistical precision are accompanied by increased parameter bias relative to the fiducial cosmology ($\Omega_m=0.31$, $\sigma_8=0.81$). As shown in Figure~\ref{fig:mcmc_results_s_range_1d}, for the range $s\in [25,85]~h^{-1}{\rm Mpc}$, the relative biases in both $\Omega_m$ and $\sigma_8$ are approximately twice as large as those obtained from the baseline $s\in[15,85]~h^{-1}{\rm Mpc}$. For the case $s\in[15,95]~h^{-1}{\rm Mpc}$, the bias in $\Omega_m$ increases by a factor of about three relative to the baseline, while the increase in the $\sigma_8$ bias is more moderate. These results highlight a trade-off between statistical constraining power and robustness to systematic shifts when incorporating larger separation scales.

For the anisotropic statistic $\widehat{W}^{\alpha}_{\Delta s}(\mu)$, the constraining power is primarily driven by the inclusion of small scales. The FoM is strongly degraded when $s_{\min}$ is increased to $25~h^{-1}{\rm Mpc}$, whereas varying $s_{\max}$ within the tested range has only a minor effect. This suggests that the anisotropic MCF gains most of its cosmological sensitivity from quasi-linear to mildly nonlinear scales, while the additional large-scale information contributes only subdominantly.

Overall, these results show that the constraining power of MCFs is mainly controlled by two factors: the inclusion of redshift-space anisotropy and the retention of small-scale information. Combining multiple marks provides an additional gain, but this gain is subdominant compared with the improvement obtained from the anisotropic statistic and from the small-scale separation bins.

\subsection{Impact of Halo Bias on Cosmological Constraints}

\begin{figure*}[htpb]
	\centering
	\begin{tabular}{cc}
		\includegraphics[width=0.48\textwidth]{"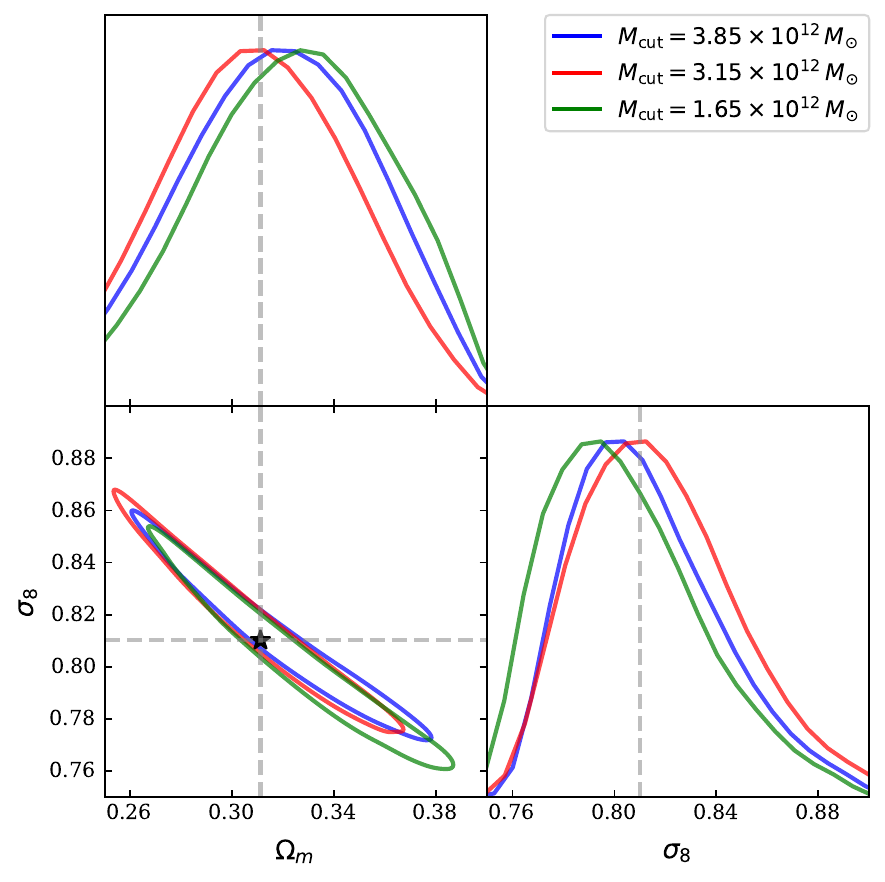"} & \includegraphics[width=0.48\textwidth]{"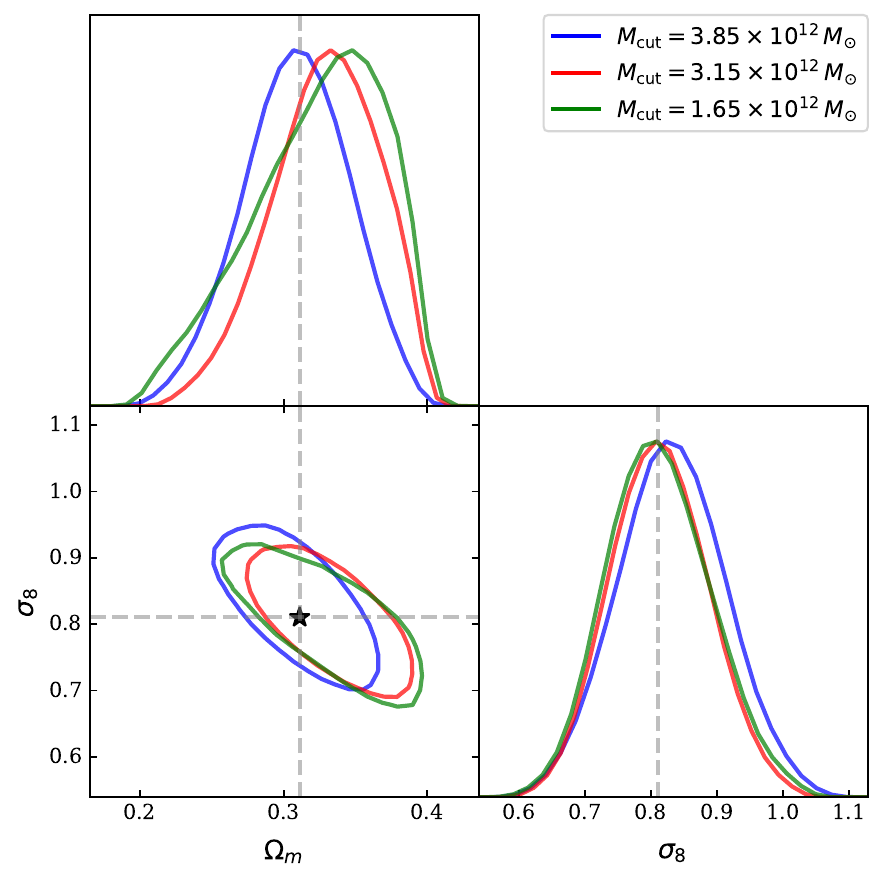"} \\
	\end{tabular}
\caption{Similar to Figure~\ref{fig:mcmc_results_alphas}, but for different halo mass cuts $M_\mathrm{cut} =3.85$, 3.15, and 1.65 $\times10^{12}~M_\odot$. The MCMC constraints for $\widehat{W}^{\alpha}(s)$ are shown in the left panel, and those for $\widehat{W}^{\alpha}_{\Delta s}(\mu)$ are shown in the right panel. In all cases, we adopt $\alpha = [-0.5, -0.3, 0, 0.3, 0.5]$ and the joint weighting scheme combining $w \propto \rho$ and $w \propto \nabla\rho/\rho$. The analysis is carried out using the fiducial binning scheme.
}
\label{fig:mcmc_results_mass_shift}
\end{figure*}

To test the robustness of our method against uncertainties in how tracers relate to the underlying matter distribution, we evaluate its performance on mock halo catalogs constructed with different halo mass selections, while keeping the emulator training data unchanged. Specifically, we impose three truncation mass thresholds, $M_\mathrm{cut} = 3.85\times10^{12}$, $3.15\times10^{12}$, and $1.65\times10^{12}~M_{\odot}$, on the original halo catalog. For each value of $M_\mathrm{cut}$, we further randomly remove ${0\%,~ 20\%,~ 60\%}$ of halos/subhalos to maintain a constant number density. This procedure isolates the impact of changes in the internal halo population (i.e., mass-dependent tracer properties) from variations in the overall clustering amplitude.

As shown in Figure~\ref{fig:mcmc_results_mass_shift}, the cosmological constraints derived from $\widehat{W}^{\alpha}(s)$ and $\widehat{W}^{\alpha}_{\Delta s}(\mu)$ remain highly consistent across all halo samples. The posterior shifts in both $\Omega_m$ and $\sigma_8$ are within $1\sigma$ of the fiducial case, indicating strong insensitivity to the adopted halo mass range. Quantitatively, relative to the fiducial case ($M_\mathrm{cut} = 3.85\times10^{12}~M_{\odot}$), the FoM of $\widehat{W}^{\alpha}(s)$ and $\widehat{W}^{\alpha}_{\Delta s}(\mu)$ changes within 10\% for $M_\mathrm{cut} = 3.15\times10^{12}$ and $1.65\times10^{12}~M_{\odot}$. This demonstrates that $\widehat{W}^{\alpha}(s)$ and $\widehat{W}^{\alpha}_{\Delta s}(\mu)$ provides stable cosmological constraints that are robust to uncertainties in the halo–mass connection, thereby alleviating a key systematic challenge in galaxy survey analyses.

\section{Conclusion}
\label{sec:conclusion}
In this work, we developed a simulation-based inference framework to test the cosmological constraining power of MCFs. We used the \textsc{Kun} simulation suite as the training set and built a Gaussian Process emulator to predict MCF statistics for different cosmological parameters. We then used the \textsc{Jiutian} simulation to estimate the covariance matrix and to validate the emulator. Finally, we applied MCMC to mock data and constrained two cosmological parameters, $\Omega_m$ and $\sigma_8$.

Our results show that MCFs can provide stronger cosmological constraints than 2PCF. This improvement is especially clear when several weighting exponents $\alpha$ are combined. For example, the joint use of $\alpha = [-0.5, -0.3, 0, 0.3, 0.5]$ gives a FoM about $2$ times larger than that from the 2PCF alone. We also find that the two weighting choices, based on the local density $\rho$ and the normalized density gradient $|\nabla \rho|/\rho$, contain complementary information. Their combination helps reduce parameter degeneracies and leads to tighter constraints.

We also studied how the results depend on scale selection. The constraining power of $\widehat{W}^{\alpha}(s)$ changes noticeably with the chosen range of $s$, showing that this statistic is sensitive to scale cuts. In comparison, the anisotropic statistic $\widehat{W}^{\alpha}_{\Delta s}(\mu)$ is less sensitive to the choice of maximum scale, although its constraining power still depends strongly on small-scale information.

An important result of this work is that the MCF constraints are stable under different halo selection choices. By changing the halo mass cut, we find that the constraints from $\widehat{W}^{\alpha}(s)$ and $\widehat{W}^{\alpha}_{\Delta s}(\mu)$ remain consistent and unbiased. This suggests that MCFs may be less sensitive to uncertainties in the galaxy-halo connection. Such robustness is important for future applications to observational galaxy surveys, where the relation between galaxies and dark matter halos is not perfectly known.

In summary, MCFs provide a useful extension of the standard 2PCF. By using environment-dependent weights such as $\rho$ and $|\nabla \rho|/\rho$, and by combining multiple values of $\alpha$, MCFs can extract additional non-Gaussian information from the cosmic density field. They improve cosmological parameter constraints and show good stability against changes in halo selection. These results suggest that MCFs are a promising and robust tool for cosmological analysis in next-generation galaxy surveys.

\begin{acknowledgements}
This work is supported by National SKA Program of China (2025SKA0160100), National Science Foundation of China (12473097), the China Manned Space Project with No. CMS-CSST-2021 (A02, A03, B01),  Guangdong Basic and Applied Basic Research Foundation (2024A1515012309), the National Natural Science Foundation of China (12373005). We also acknowledge the Beijing Super Cloud Center (BSCC) and Beijing Beilong Super Cloud Computing Co., Ltd (\url{http://www.blsc.cn/}) for providing HPC resources that have significantly contributed to the research results presented in this paper.
\end{acknowledgements}

\bibliography{apssamp}

@ARTICLE{gong2019csst,
       author = {{Gong}, Yan and {Liu}, Xiangkun and {Cao}, Ye and {Chen}, Xuelei and {Fan}, Zuhui and {Li}, Ran and {Li}, Xiao-Dong and {Li}, Zhigang and {Zhang}, Xin and {Zhan}, Hu},
        title = "{Cosmology from the Chinese Space Station Optical Survey (CSS-OS)}",
      journal = {\apj},
         year = 2019,
        month = oct,
       volume = {883},
       number = {2},
          eid = {203},
        pages = {203},
          doi = {10.3847/1538-4357/ab391e},
archivePrefix = {arXiv},
       eprint = {1901.04634},
 primaryClass = {astro-ph.CO},
       adsurl = {https://ui.adsabs.harvard.edu/abs/2019ApJ...883..203G}
}

@ARTICLE{eucild2011,
       author = {{Laureijs}, R. and {Amiaux}, J. and {Arduini}, S. and {Augu{\`e}res}, J. -L. and {Brinchmann}, J. and {Cole}, R. and {Cropper}, M. and {Dabin}, C. and {Duvet}, L. and {Ealet}, A. and {Garilli}, B. and {Gondoin}, P. and {Guzzo}, L. and {Hoar}, J. and {Hoekstra}, H. and {Holmes}, R. and {Kitching}, T. and {Maciaszek}, T. and {Mellier}, Y. and {Pasian}, F. and {Percival}, W. and {Rhodes}, J. and {Saavedra Criado}, G. and {Sauvage}, M. and {Scaramella}, R. and {Valenziano}, L. and {Warren}, S. and {Bender}, R. and {Castander}, F. and {Cimatti}, A. and {Le F{\`e}vre}, O. and {Kurki-Suonio}, H. and {Levi}, M. and {Lilje}, P. and {Meylan}, G. and {Nichol}, R. and {Pedersen}, K. and {Popa}, V. and {Rebolo Lopez}, R. and {Rix}, H. -W. and {Rottgering}, H. and {Zeilinger}, W. and {Grupp}, F. and {Hudelot}, P. and {Massey}, R. and {Meneghetti}, M. and {Miller}, L. and {Paltani}, S. and {Paulin-Henriksson}, S. and {Pires}, S. and {Saxton}, C. and {Schrabback}, T. and {Seidel}, G. and {Walsh}, J. and {Aghanim}, N. and {Amendola}, L. and {Bartlett}, J. and {Baccigalupi}, C. and {Beaulieu}, J. -P. and {Benabed}, K. and {Cuby}, J. -G. and {Elbaz}, D. and {Fosalba}, P. and {Gavazzi}, G. and {Helmi}, A. and {Hook}, I. and {Irwin}, M. and {Kneib}, J. -P. and {Kunz}, M. and {Mannucci}, F. and {Moscardini}, L. and {Tao}, C. and {Teyssier}, R. and {Weller}, J. and {Zamorani}, G. and {Zapatero Osorio}, M.~R. and {Boulade}, O. and {Foumond}, J.~J. and {Di Giorgio}, A. and {Guttridge}, P. and {James}, A. and {Kemp}, M. and {Martignac}, J. and {Spencer}, A. and {Walton}, D. and {Bl{\"u}mchen}, T. and {Bonoli}, C. and {Bortoletto}, F. and {Cerna}, C. and {Corcione}, L. and {Fabron}, C. and {Jahnke}, K. and {Ligori}, S. and {Madrid}, F. and {Martin}, L. and {Morgante}, G. and {Pamplona}, T. and {Prieto}, E. and {Riva}, M. and {Toledo}, R. and {Trifoglio}, M. and {Zerbi}, F. and {Abdalla}, F. and {Douspis}, M. and {Grenet}, C. and {Borgani}, S. and {Bouwens}, R. and {Courbin}, F. and {Delouis}, J. -M. and {Dubath}, P. and {Fontana}, A. and {Frailis}, M. and {Grazian}, A. and {Koppenh{\"o}fer}, J. and {Mansutti}, O. and {Melchior}, M. and {Mignoli}, M. and {Mohr}, J. and {Neissner}, C. and {Noddle}, K. and {Poncet}, M. and {Scodeggio}, M. and {Serrano}, S. and {Shane}, N. and {Starck}, J. -L. and {Surace}, C. and {Taylor}, A. and {Verdoes-Kleijn}, G. and {Vuerli}, C. and {Williams}, O.~R. and {Zacchei}, A. and {Altieri}, B. and {Escudero Sanz}, I. and {Kohley}, R. and {Oosterbroek}, T. and {Astier}, P. and {Bacon}, D. and {Bardelli}, S. and {Baugh}, C. and {Bellagamba}, F. and {Benoist}, C. and {Bianchi}, D. and {Biviano}, A. and {Branchini}, E. and {Carbone}, C. and {Cardone}, V. and {Clements}, D. and {Colombi}, S. and {Conselice}, C. and {Cresci}, G. and {Deacon}, N. and {Dunlop}, J. and {Fedeli}, C. and {Fontanot}, F. and {Franzetti}, P. and {Giocoli}, C. and {Garcia-Bellido}, J. and {Gow}, J. and {Heavens}, A. and {Hewett}, P. and {Heymans}, C. and {Holland}, A. and {Huang}, Z. and {Ilbert}, O. and {Joachimi}, B. and {Jennins}, E. and {Kerins}, E. and {Kiessling}, A. and {Kirk}, D. and {Kotak}, R. and {Krause}, O. and {Lahav}, O. and {van Leeuwen}, F. and {Lesgourgues}, J. and {Lombardi}, M. and {Magliocchetti}, M. and {Maguire}, K. and {Majerotto}, E. and {Maoli}, R. and {Marulli}, F. and {Maurogordato}, S. and {McCracken}, H. and {McLure}, R. and {Melchiorri}, A. and {Merson}, A. and {Moresco}, M. and {Nonino}, M. and {Norberg}, P. and {Peacock}, J. and {Pello}, R. and {Penny}, M. and {Pettorino}, V. and {Di Porto}, C. and {Pozzetti}, L. and {Quercellini}, C. and {Radovich}, M. and {Rassat}, A. and {Roche}, N. and {Ronayette}, S. and {Rossetti}, E.},
        title = "{Euclid Definition Study Report}",
      journal = {arXiv e-prints},
         year = 2011,
        month = oct,
          eid = {arXiv:1110.3193},
        pages = {arXiv:1110.3193},
          doi = {10.48550/arXiv.1110.3193},
archivePrefix = {arXiv},
       eprint = {1110.3193},
 primaryClass = {astro-ph.CO},
       adsurl = {https://ui.adsabs.harvard.edu/abs/2011arXiv1110.3193L}
}

@ARTICLE{lsst2009,
       author = {{LSST Science Collaboration} and {Abell}, Paul A. and {Allison}, Julius and {Anderson}, Scott F. and {Andrew}, John R. and {Angel}, J. Roger P. and {Armus}, Lee and {Arnett}, David and {Asztalos}, S.~J. and {Axelrod}, Tim S. and {Bailey}, Stephen and {Ballantyne}, D.~R. and {Bankert}, Justin R. and {Barkhouse}, Wayne A. and {Barr}, Jeffrey D. and {Barrientos}, L. Felipe and {Barth}, Aaron J. and {Bartlett}, James G. and {Becker}, Andrew C. and {Becla}, Jacek and {Beers}, Timothy C. and {Bernstein}, Joseph P. and {Biswas}, Rahul and {Blanton}, Michael R. and {Bloom}, Joshua S. and {Bochanski}, John J. and {Boeshaar}, Pat and {Borne}, Kirk D. and {Bradac}, Marusa and {Brandt}, W.~N. and {Bridge}, Carrie R. and {Brown}, Michael E. and {Brunner}, Robert J. and {Bullock}, James S. and {Burgasser}, Adam J. and {Burge}, James H. and {Burke}, David L. and {Cargile}, Phillip A. and {Chandrasekharan}, Srinivasan and {Chartas}, George and {Chesley}, Steven R. and {Chu}, You-Hua and {Cinabro}, David and {Claire}, Mark W. and {Claver}, Charles F. and {Clowe}, Douglas and {Connolly}, A.~J. and {Cook}, Kem H. and {Cooke}, Jeff and {Cooray}, Asantha and {Covey}, Kevin R. and {Culliton}, Christopher S. and {de Jong}, Roelof and {de Vries}, Willem H. and {Debattista}, Victor P. and {Delgado}, Francisco and {Dell'Antonio}, Ian P. and {Dhital}, Saurav and {Di Stefano}, Rosanne and {Dickinson}, Mark and {Dilday}, Benjamin and {Djorgovski}, S.~G. and {Dobler}, Gregory and {Donalek}, Ciro and {Dubois-Felsmann}, Gregory and {Durech}, Josef and {Eliasdottir}, Ardis and {Eracleous}, Michael and {Eyer}, Laurent and {Falco}, Emilio E. and {Fan}, Xiaohui and {Fassnacht}, Christopher D. and {Ferguson}, Harry C. and {Fernandez}, Yanga R. and {Fields}, Brian D. and {Finkbeiner}, Douglas and {Figueroa}, Eduardo E. and {Fox}, Derek B. and {Francke}, Harold and {Frank}, James S. and {Frieman}, Josh and {Fromenteau}, Sebastien and {Furqan}, Muhammad and {Galaz}, Gaspar and {Gal-Yam}, A. and {Garnavich}, Peter and {Gawiser}, Eric and {Geary}, John and {Gee}, Perry and {Gibson}, Robert R. and {Gilmore}, Kirk and {Grace}, Emily A. and {Green}, Richard F. and {Gressler}, William J. and {Grillmair}, Carl J. and {Habib}, Salman and {Haggerty}, J.~S. and {Hamuy}, Mario and {Harris}, Alan W. and {Hawley}, Suzanne L. and {Heavens}, Alan F. and {Hebb}, Leslie and {Henry}, Todd J. and {Hileman}, Edward and {Hilton}, Eric J. and {Hoadley}, Keri and {Holberg}, J.~B. and {Holman}, Matt J. and {Howell}, Steve B. and {Infante}, Leopoldo and {Ivezic}, Zeljko and {Jacoby}, Suzanne H. and {Jain}, Bhuvnesh and {R} and {Jedicke} and {Jee}, M. James and {Garrett Jernigan}, J. and {Jha}, Saurabh W. and {Johnston}, Kathryn V. and {Jones}, R. Lynne and {Juric}, Mario and {Kaasalainen}, Mikko and {Styliani} and {Kafka} and {Kahn}, Steven M. and {Kaib}, Nathan A. and {Kalirai}, Jason and {Kantor}, Jeff and {Kasliwal}, Mansi M. and {Keeton}, Charles R. and {Kessler}, Richard and {Knezevic}, Zoran and {Kowalski}, Adam and {Krabbendam}, Victor L. and {Krughoff}, K. Simon and {Kulkarni}, Shrinivas and {Kuhlman}, Stephen and {Lacy}, Mark and {Lepine}, Sebastien and {Liang}, Ming and {Lien}, Amy and {Lira}, Paulina and {Long}, Knox S. and {Lorenz}, Suzanne and {Lotz}, Jennifer M. and {Lupton}, R.~H. and {Lutz}, Julie and {Macri}, Lucas M. and {Mahabal}, Ashish A. and {Mandelbaum}, Rachel and {Marshall}, Phil and {May}, Morgan and {McGehee}, Peregrine M. and {Meadows}, Brian T. and {Meert}, Alan and {Milani}, Andrea and {Miller}, Christopher J. and {Miller}, Michelle and {Mills}, David and {Minniti}, Dante and {Monet}, David and {Mukadam}, Anjum S. and {Nakar}, Ehud and {Neill}, Douglas R. and {Newman}, Jeffrey A. and {Nikolaev}, Sergei and {Nordby}, Martin and {O'Connor}, Paul and {Oguri}, Masamune and {Oliver}, John and {Olivier}, Scot S. and {Olsen}, Julia K. and {Olsen}, Knut and {Olszewski}, Edward W. and {Oluseyi}, Hakeem and {Padilla}, Nelson D. and {Parker}, Alex and {Pepper}, Joshua and {Peterson}, John R. and {Petry}, Catherine and {Pinto}, Philip A. and {Pizagno}, James L. and {Popescu}, Bogdan and {Prsa}, Andrej and {Radcka}, Veljko and {Raddick}, M. Jordan and {Rasmussen}, Andrew and {Rau}, Arne and {Rho}, Jeonghee and {Rhoads}, James E. and {Richards}, Gordon T. and {Ridgway}, Stephen T. and {Robertson}, Brant E. and {Roskar}, Rok and {Saha}, Abhijit and {Sarajedini}, Ata and {Scannapieco}, Evan and {Schalk}, Terry and {Schindler}, Rafe and {Schmidt}, Samuel},
        title = "{LSST Science Book, Version 2.0}",
      journal = {arXiv e-prints},
         year = 2009,
        month = dec,
          eid = {arXiv:0912.0201},
        pages = {arXiv:0912.0201},
          doi = {10.48550/arXiv.0912.0201},
archivePrefix = {arXiv},
       eprint = {0912.0201},
 primaryClass = {astro-ph.IM},
       adsurl = {https://ui.adsabs.harvard.edu/abs/2009arXiv0912.0201L}
}

@ARTICLE{desi2016,
       author = {{DESI Collaboration} and {Aghamousa}, Amir and {Aguilar}, Jessica and {Ahlen}, Steve and {Alam}, Shadab and {Allen}, Lori E. and {Allende Prieto}, Carlos and {Annis}, James and {Bailey}, Stephen and {Balland}, Christophe and {Ballester}, Otger and {Baltay}, Charles and {Beaufore}, Lucas and {Bebek}, Chris and {Beers}, Timothy C. and {Bell}, Eric F. and {Bernal}, Jos{\'e} Luis and {Besuner}, Robert and {Beutler}, Florian and {Blake}, Chris and {Bleuler}, Hannes and {Blomqvist}, Michael and {Blum}, Robert and {Bolton}, Adam S. and {Briceno}, Cesar and {Brooks}, David and {Brownstein}, Joel R. and {Buckley-Geer}, Elizabeth and {Burden}, Angela and {Burtin}, Etienne and {Busca}, Nicolas G. and {Cahn}, Robert N. and {Cai}, Yan-Chuan and {Cardiel-Sas}, Laia and {Carlberg}, Raymond G. and {Carton}, Pierre-Henri and {Casas}, Ricard and {Castander}, Francisco J. and {Cervantes-Cota}, Jorge L. and {Claybaugh}, Todd M. and {Close}, Madeline and {Coker}, Carl T. and {Cole}, Shaun and {Comparat}, Johan and {Cooper}, Andrew P. and {Cousinou}, M. -C. and {Crocce}, Martin and {Cuby}, Jean-Gabriel and {Cunningham}, Daniel P. and {Davis}, Tamara M. and {Dawson}, Kyle S. and {de la Macorra}, Axel and {De Vicente}, Juan and {Delubac}, Timoth{\'e}e and {Derwent}, Mark and {Dey}, Arjun and {Dhungana}, Govinda and {Ding}, Zhejie and {Doel}, Peter and {Duan}, Yutong T. and {Ealet}, Anne and {Edelstein}, Jerry and {Eftekharzadeh}, Sarah and {Eisenstein}, Daniel J. and {Elliott}, Ann and {Escoffier}, St{\'e}phanie and {Evatt}, Matthew and {Fagrelius}, Parker and {Fan}, Xiaohui and {Fanning}, Kevin and {Farahi}, Arya and {Farihi}, Jay and {Favole}, Ginevra and {Feng}, Yu and {Fernandez}, Enrique and {Findlay}, Joseph R. and {Finkbeiner}, Douglas P. and {Fitzpatrick}, Michael J. and {Flaugher}, Brenna and {Flender}, Samuel and {Font-Ribera}, Andreu and {Forero-Romero}, Jaime E. and {Fosalba}, Pablo and {Frenk}, Carlos S. and {Fumagalli}, Michele and {Gaensicke}, Boris T. and {Gallo}, Giuseppe and {Garcia-Bellido}, Juan and {Gaztanaga}, Enrique and {Pietro Gentile Fusillo}, Nicola and {Gerard}, Terry and {Gershkovich}, Irena and {Giannantonio}, Tommaso and {Gillet}, Denis and {Gonzalez-de-Rivera}, Guillermo and {Gonzalez-Perez}, Violeta and {Gott}, Shelby and {Graur}, Or and {Gutierrez}, Gaston and {Guy}, Julien and {Habib}, Salman and {Heetderks}, Henry and {Heetderks}, Ian and {Heitmann}, Katrin and {Hellwing}, Wojciech A. and {Herrera}, David A. and {Ho}, Shirley and {Holland}, Stephen and {Honscheid}, Klaus and {Huff}, Eric and {Hutchinson}, Timothy A. and {Huterer}, Dragan and {Hwang}, Ho Seong and {Illa Laguna}, Joseph Maria and {Ishikawa}, Yuzo and {Jacobs}, Dianna and {Jeffrey}, Niall and {Jelinsky}, Patrick and {Jennings}, Elise and {Jiang}, Linhua and {Jimenez}, Jorge and {Johnson}, Jennifer and {Joyce}, Richard and {Jullo}, Eric and {Juneau}, St{\'e}phanie and {Kama}, Sami and {Karcher}, Armin and {Karkar}, Sonia and {Kehoe}, Robert and {Kennamer}, Noble and {Kent}, Stephen and {Kilbinger}, Martin and {Kim}, Alex G. and {Kirkby}, David and {Kisner}, Theodore and {Kitanidis}, Ellie and {Kneib}, Jean-Paul and {Koposov}, Sergey and {Kovacs}, Eve and {Koyama}, Kazuya and {Kremin}, Anthony and {Kron}, Richard and {Kronig}, Luzius and {Kueter-Young}, Andrea and {Lacey}, Cedric G. and {Lafever}, Robin and {Lahav}, Ofer and {Lambert}, Andrew and {Lampton}, Michael and {Landriau}, Martin and {Lang}, Dustin and {Lauer}, Tod R. and {Le Goff}, Jean-Marc and {Le Guillou}, Laurent and {Le Van Suu}, Auguste and {Lee}, Jae Hyeon and {Lee}, Su-Jeong and {Leitner}, Daniela and {Lesser}, Michael and {Levi}, Michael E. and {L'Huillier}, Benjamin and {Li}, Baojiu and {Liang}, Ming and {Lin}, Huan and {Linder}, Eric and {Loebman}, Sarah R. and {Luki{\'c}}, Zarija and {Ma}, Jun and {MacCrann}, Niall and {Magneville}, Christophe and {Makarem}, Laleh and {Manera}, Marc and {Manser}, Christopher J. and {Marshall}, Robert and {Martini}, Paul and {Massey}, Richard and {Matheson}, Thomas and {McCauley}, Jeremy and {McDonald}, Patrick and {McGreer}, Ian D. and {Meisner}, Aaron and {Metcalfe}, Nigel and {Miller}, Timothy N. and {Miquel}, Ramon and {Moustakas}, John and {Myers}, Adam and {Naik}, Milind and {Newman}, Jeffrey A. and {Nichol}, Robert C. and {Nicola}, Andrina and {Nicolati da Costa}, Luiz and {Nie}, Jundan and {Niz}, Gustavo and {Norberg}, Peder and {Nord}, Brian and {Norman}, Dara and {Nugent}, Peter and {O'Brien}, Thomas and {Oh}, Minji and {Olsen}, Knut A.~G.},
        title = "{The DESI Experiment Part I: Science,Targeting, and Survey Design}",
      journal = {arXiv e-prints},
         year = 2016,
        month = oct,
          eid = {arXiv:1611.00036},
        pages = {arXiv:1611.00036},
          doi = {10.48550/arXiv.1611.00036},
archivePrefix = {arXiv},
       eprint = {1611.00036},
 primaryClass = {astro-ph.IM},
       adsurl = {https://ui.adsabs.harvard.edu/abs/2016arXiv161100036D}
}

@ARTICLE{eucild2024,
       author = {{Euclid Collaboration} and {Mellier}, Y. and {Abdurro'uf} and {Acevedo Barroso}, J.~A. and {Ach{\'u}carro}, A. and {Adamek}, J. and {Adam}, R. and {Addison}, G.~E. and {Aghanim}, N. and {Aguena}, M. and {Ajani}, V. and {Akrami}, Y. and {Al-Bahlawan}, A. and {Alavi}, A. and {Albuquerque}, I.~S. and {Alestas}, G. and {Alguero}, G. and {Allaoui}, A. and {Allen}, S.~W. and {Allevato}, V. and {Alonso-Tetilla}, A.~V. and {Altieri}, B. and {Alvarez-Candal}, A. and {Alvi}, S. and {Amara}, A. and {Amendola}, L. and {Amiaux}, J. and {Andika}, I.~T. and {Andreon}, S. and {Andrews}, A. and {Angora}, G. and {Angulo}, R.~E. and {Annibali}, F. and {Anselmi}, A. and {Anselmi}, S. and {Arcari}, S. and {Archidiacono}, M. and {Aric{\`o}}, G. and {Arnaud}, M. and {Arnouts}, S. and {Asgari}, M. and {Asorey}, J. and {Atayde}, L. and {Atek}, H. and {Atrio-Barandela}, F. and {Aubert}, M. and {Aubourg}, E. and {Auphan}, T. and {Auricchio}, N. and {Aussel}, B. and {Aussel}, H. and {Avelino}, P.~P. and {Avgoustidis}, A. and {Avila}, S. and {Awan}, S. and {Azzollini}, R. and {Baccigalupi}, C. and {Bachelet}, E. and {Bacon}, D. and {Baes}, M. and {Bagley}, M.~B. and {Bahr-Kalus}, B. and {Balaguera-Antolinez}, A. and {Balbinot}, E. and {Balcells}, M. and {Baldi}, M. and {Baldry}, I. and {Balestra}, A. and {Ballardini}, M. and {Ballester}, O. and {Balogh}, M. and {Ba{\~n}ados}, E. and {Barbier}, R. and {Bardelli}, S. and {Baron}, M. and {Barreiro}, T. and {Barrena}, R. and {Barriere}, J. -C. and {Barros}, B.~J. and {Barthelemy}, A. and {Bartolo}, N. and {Basset}, A. and {Battaglia}, P. and {Battisti}, A.~J. and {Baugh}, C.~M. and {Baumont}, L. and {Bazzanini}, L. and {Beaulieu}, J. -P. and {Beckmann}, V. and {Belikov}, A.~N. and {Bel}, J. and {Bellagamba}, F. and {Bella}, M. and {Bellini}, E. and {Benabed}, K. and {Bender}, R. and {Benevento}, G. and {Bennett}, C.~L. and {Benson}, K. and {Bergamini}, P. and {Bermejo-Climent}, J.~R. and {Bernardeau}, F. and {Bertacca}, D. and {Berthe}, M. and {Berthier}, J. and {Bethermin}, M. and {Beutler}, F. and {Bevillon}, C. and {Bhargava}, S. and {Bhatawdekar}, R. and {Bianchi}, D. and {Bisigello}, L. and {Biviano}, A. and {Blake}, R.~P. and {Blanchard}, A. and {Blazek}, J. and {Blot}, L. and {Bosco}, A. and {Bodendorf}, C. and {Boenke}, T. and {B{\"o}hringer}, H. and {Boldrini}, P. and {Bolzonella}, M. and {Bonchi}, A. and {Bonici}, M. and {Bonino}, D. and {Bonino}, L. and {Bonvin}, C. and {Bon}, W. and {Booth}, J.~T. and {Borgani}, S. and {Borlaff}, A.~S. and {Borsato}, E. and {Bosco}, A. and {Bose}, B. and {Botticella}, M.~T. and {Boucaud}, A. and {Bouche}, F. and {Boucher}, J.~S. and {Boutigny}, D. and {Bouvard}, T. and {Bouwens}, R. and {Bouy}, H. and {Bowler}, R.~A.~A. and {Bozza}, V. and {Bozzo}, E. and {Branchini}, E. and {Brando}, G. and {Brau-Nogue}, S. and {Brekke}, P. and {Bremer}, M.~N. and {Brescia}, M. and {Breton}, M. -A. and {Brinchmann}, J. and {Brinckmann}, T. and {Brockley-Blatt}, C. and {Brodwin}, M. and {Brouard}, L. and {Brown}, M.~L. and {Bruton}, S. and {Bucko}, J. and {Buddelmeijer}, H. and {Buenadicha}, G. and {Buitrago}, F. and {Burger}, P. and {Burigana}, C. and {Busillo}, V. and {Busonero}, D. and {Cabanac}, R. and {Cabayol-Garcia}, L. and {Cagliari}, M.~S. and {Caillat}, A. and {Caillat}, L. and {Calabrese}, M. and {Calabro}, A. and {Calderone}, G. and {Calura}, F. and {Camacho Quevedo}, B. and {Camera}, S. and {Campos}, L. and {Canas-Herrera}, G. and {Candini}, G.~P. and {Cantiello}, M. and {Capobianco}, V. and {Cappellaro}, E. and {Cappelluti}, N. and {Cappi}, A. and {Caputi}, K.~I. and {Cara}, C. and {Carbone}, C. and {Cardone}, V.~F. and {Carella}, E. and {Carlberg}, R.~G. and {Carle}, M. and {Carminati}, L. and {Caro}, F. and {Carrasco}, J.~M. and {Carretero}, J. and {Carrilho}, P. and {Carron Duque}, J.},
        title = "{Euclid. I. Overview of the Euclid mission}",
      journal = {arXiv e-prints},
         year = 2024,
        month = may,
          eid = {arXiv:2405.13491},
        pages = {arXiv:2405.13491},
          doi = {10.48550/arXiv.2405.13491},
archivePrefix = {arXiv},
       eprint = {2405.13491},
 primaryClass = {astro-ph.CO},
       adsurl = {https://ui.adsabs.harvard.edu/abs/2024arXiv240513491E}
}

@ARTICLE{rst2019,
       author = {{Dore}, Olivier and {Hirata}, C. and {Wang}, Y. and {Weinberg}, D. and {Eifler}, Tim and {Foley}, R.~J. and {Heinrich}, C. He and {Krause}, E. and {Perlmutter}, S. and {Pisani}, A. and {Scolnic}, D. and {Spergel}, D.~N. and {Suntzeff}, N. and {Aldering}, G. and {Baltay}, C. and {Capak}, P. and {Choi}, A. and {Dvorkin}, C. and {Fall}, S.~M. and {Fang}, X. and {Fruchter}, A. and {Galbany}, L. and {Ho}, S. and {Hounsell}, R. and {Izard}, A. and {Jain}, B. and {Koekemoer}, A.~M. and {Kruk}, J. and {Leauthaud}, A. and {Malhotra}, S. and {Mandelbaum}, R. and {Massara}, E. and {Masters}, D. and {Miyatake}, H. and {Plazas}, A. and {Rhoads}, J. and {Rhodes}, J. and {Rose}, B. and {Rubin}, D. and {Sako}, M. and {Samushia}, L. and {Shirasaki}, M. and {Simet}, M. and {Takada}, M. and {Troxel}, M.~A. and {Wu}, H. and {Yoshida}, N. and {Zhai}, Z.},
        title = "{WFIRST: The Essential Cosmology Space Observatory for the Coming Decade}",
      journal = {\baas},
         year = 2019,
        month = may,
       volume = {51},
       number = {3},
          eid = {341},
        pages = {341},
          doi = {10.48550/arXiv.1904.01174},
archivePrefix = {arXiv},
       eprint = {1904.01174},
 primaryClass = {astro-ph.CO},
       adsurl = {https://ui.adsabs.harvard.edu/abs/2019BAAS...51c.341D}
}

@ARTICLE{2df:Colless:2003wz,
   author = {{Colless}, M. and {Peterson}, B.~A. and {Jackson}, C. and {Peacock}, J.~A. and {Cole}, S. and {Norberg}, P. and {Baldry}, I.~K. and {Baugh}, C.~M. and {Bland-Hawthorn}, J. and {Bridges}, T. and {Cannon}, R. and {Collins}, C. and {Couch}, W. and {Cross}, N. and {Dalton}, G. and {De Propris}, R. and {Driver}, S.~P. and {Efstathiou}, G. and {Ellis}, R.~S. and {Frenk}, C.~S. and {Glazebrook}, K. and {Lahav}, O. and {Lewis}, I. and {Lumsden}, S. and {Maddox}, S. and {Madgwick}, D. and {Sutherland}, W. and {Taylor}, K.},
    title = "{The 2dF Galaxy Redshift Survey: Final Data Release}",
  journal = {arXiv Astrophysics e-prints},
   eprint = {astro-ph/0306581},
     year = 2003,
    month = jun,
   adsurl = {https://ui.adsabs.harvard.edu/abs/2003astro.ph..6581C}
}

@ARTICLE{beutler20116df,
   author = {{Beutler}, F. and {Blake}, C. and {Colless}, M. and {Jones}, D.~H. and 
	{Staveley-Smith}, L. and {Poole}, G.~B. and {Campbell}, L. and 
	{Parker}, Q. and {Saunders}, W. and {Watson}, F.},
    title = "{The 6dF Galaxy Survey: z{\ap} 0 measurements of the growth rate and {$\sigma$}$_{8}$}",
  journal = {mnras},
archivePrefix = "arXiv",
   eprint = {1204.4725},
     year = 2012,
    month = jul,
   volume = 423,
    pages = {3430-3444},
      doi = {10.1111/j.1365-2966.2012.21136.x},
   adsurl = {https://ui.adsabs.harvard.edu/abs/2012MNRAS.423.3430B}
}

@ARTICLE{blake2011wigglez,
   author = {{Blake}, C. and {Brough}, S. and {Colless}, M. and {Contreras}, C. and 
	{Couch}, W. and {Croom}, S. and {Davis}, T. and {Drinkwater}, M.~J. and 
	{Forster}, K. and {Gilbank}, D. and {Gladders}, M. and {Glazebrook}, K. and 
	{Jelliffe}, B. and {Jurek}, R.~J. and {Li}, I.-H. and {Madore}, B. and 
	{Martin}, D.~C. and {Pimbblet}, K. and {Poole}, G.~B. and {Pracy}, M. and 
	{Sharp}, R. and {Wisnioski}, E. and {Woods}, D. and {Wyder}, T.~K. and 
	{Yee}, H.~K.~C.},
    title = "{The WiggleZ Dark Energy Survey: the growth rate of cosmic structure since redshift z=0.9}",
  journal = {mnras},
archivePrefix = "arXiv",
   eprint = {1104.2948},
     year = 2011,
    month = aug,
   volume = 415,
    pages = {2876-2891},
      doi = {10.1111/j.1365-2966.2011.18903.x},
   adsurl = {https://ui.adsabs.harvard.edu/abs/2011MNRAS.415.2876B}
}

@ARTICLE{blake2011wigglezb,
   author = {{Blake}, C. and {Glazebrook}, K. and {Davis}, T.~M. and {Brough}, S. and 
	{Colless}, M. and {Contreras}, C. and {Couch}, W. and {Croom}, S. and 
	{Drinkwater}, M.~J. and {Forster}, K. and {Gilbank}, D. and 
	{Gladders}, M. and {Jelliffe}, B. and {Jurek}, R.~J. and {Li}, I.-H. and 
	{Madore}, B. and {Martin}, D.~C. and {Pimbblet}, K. and {Poole}, G.~B. and 
	{Pracy}, M. and {Sharp}, R. and {Wisnioski}, E. and {Woods}, D. and 
	{Wyder}, T.~K. and {Yee}, H.~K.~C.},
    title = "{The WiggleZ Dark Energy Survey: measuring the cosmic expansion history using the Alcock-Paczynski test and distant supernovae}",
  journal = {mnras},
archivePrefix = "arXiv",
   eprint = {1108.2637},
     year = 2011,
    month = dec,
   volume = 418,
    pages = {1725-1735},
      doi = {10.1111/j.1365-2966.2011.19606.x},
   adsurl = {https://ui.adsabs.harvard.edu/abs/2011MNRAS.418.1725B}
}

@ARTICLE{york2000sloan,
   author = {{York}, D.~G. and {Adelman}, J. and {Anderson}, Jr., J.~E. and 
	{Anderson}, S.~F. and {Annis}, J. and {Bahcall}, N.~A. and {Bakken}, J.~A. and 
	{Barkhouser}, R. and {Bastian}, S. and {Berman}, E. and {Boroski}, W.~N. and 
	{Bracker}, S. and {Briegel}, C. and {Briggs}, J.~W. and {Brinkmann}, J. and 
	{Brunner}, R. and {Burles}, S. and {Carey}, L. and {Carr}, M.~A. and 
	{Castander}, F.~J. and {Chen}, B. and {Colestock}, P.~L. and 
	{Connolly}, A.~J. and {Crocker}, J.~H. and {Csabai}, I. and 
	{Czarapata}, P.~C. and {Davis}, J.~E. and {Doi}, M. and {Dombeck}, T. and 
	{Eisenstein}, D. and {Ellman}, N. and {Elms}, B.~R. and {Evans}, M.~L. and 
	{Fan}, X. and {Federwitz}, G.~R. and {Fiscelli}, L. and {Friedman}, S. and 
	{Frieman}, J.~A. and {Fukugita}, M. and {Gillespie}, B. and 
	{Gunn}, J.~E. and {Gurbani}, V.~K. and {de Haas}, E. and {Haldeman}, M. and 
	{Harris}, F.~H. and {Hayes}, J. and {Heckman}, T.~M. and {Hennessy}, G.~S. and 
	{Hindsley}, R.~B. and {Holm}, S. and {Holmgren}, D.~J. and {Huang}, C.-h. and 
	{Hull}, C. and {Husby}, D. and {Ichikawa}, S.-I. and {Ichikawa}, T. and 
	{Ivezi{\'c}}, {\v Z}. and {Kent}, S. and {Kim}, R.~S.~J. and 
	{Kinney}, E. and {Klaene}, M. and {Kleinman}, A.~N. and {Kleinman}, S. and 
	{Knapp}, G.~R. and {Korienek}, J. and {Kron}, R.~G. and {Kunszt}, P.~Z. and 
	{Lamb}, D.~Q. and {Lee}, B. and {Leger}, R.~F. and {Limmongkol}, S. and 
	{Lindenmeyer}, C. and {Long}, D.~C. and {Loomis}, C. and {Loveday}, J. and 
	{Lucinio}, R. and {Lupton}, R.~H. and {MacKinnon}, B. and {Mannery}, E.~J. and 
	{Mantsch}, P.~M. and {Margon}, B. and {McGehee}, P. and {McKay}, T.~A. and 
	{Meiksin}, A. and {Merelli}, A. and {Monet}, D.~G. and {Munn}, J.~A. and 
	{Narayanan}, V.~K. and {Nash}, T. and {Neilsen}, E. and {Neswold}, R. and 
	{Newberg}, H.~J. and {Nichol}, R.~C. and {Nicinski}, T. and 
	{Nonino}, M. and {Okada}, N. and {Okamura}, S. and {Ostriker}, J.~P. and 
	{Owen}, R. and {Pauls}, A.~G. and {Peoples}, J. and {Peterson}, R.~L. and 
	{Petravick}, D. and {Pier}, J.~R. and {Pope}, A. and {Pordes}, R. and 
	{Prosapio}, A. and {Rechenmacher}, R. and {Quinn}, T.~R. and 
	{Richards}, G.~T. and {Richmond}, M.~W. and {Rivetta}, C.~H. and 
	{Rockosi}, C.~M. and {Ruthmansdorfer}, K. and {Sandford}, D. and 
	{Schlegel}, D.~J. and {Schneider}, D.~P. and {Sekiguchi}, M. and 
	{Sergey}, G. and {Shimasaku}, K. and {Siegmund}, W.~A. and {Smee}, S. and 
	{Smith}, J.~A. and {Snedden}, S. and {Stone}, R. and {Stoughton}, C. and 
	{Strauss}, M.~A. and {Stubbs}, C. and {SubbaRao}, M. and {Szalay}, A.~S. and 
	{Szapudi}, I. and {Szokoly}, G.~P. and {Thakar}, A.~R. and {Tremonti}, C. and 
	{Tucker}, D.~L. and {Uomoto}, A. and {Vanden Berk}, D. and {Vogeley}, M.~S. and 
	{Waddell}, P. and {Wang}, S.-i. and {Watanabe}, M. and {Weinberg}, D.~H. and 
	{Yanny}, B. and {Yasuda}, N. and {SDSS Collaboration}},
    title = "{The Sloan Digital Sky Survey: Technical Summary}",
  journal = {aj},
   eprint = {astro-ph/0006396},
     year = 2000,
    month = sep,
   volume = 120,
    pages = {1579-1587},
      doi = {10.1086/301513},
   adsurl = {https://ui.adsabs.harvard.edu/abs/2000AJ....120.1579Y}
}

@ARTICLE{Eisenstein:2005su,
   author = {{Eisenstein}, D.~J. and {Zehavi}, I. and {Hogg}, D.~W. and {Scoccimarro}, R. and 
	{Blanton}, M.~R. and {Nichol}, R.~C. and {Scranton}, R. and 
	{Seo}, H.-J. and {Tegmark}, M. and {Zheng}, Z. and {Anderson}, S.~F. and 
	{Annis}, J. and {Bahcall}, N. and {Brinkmann}, J. and {Burles}, S. and 
	{Castander}, F.~J. and {Connolly}, A. and {Csabai}, I. and {Doi}, M. and 
	{Fukugita}, M. and {Frieman}, J.~A. and {Glazebrook}, K. and 
	{Gunn}, J.~E. and {Hendry}, J.~S. and {Hennessy}, G. and {Ivezi{\'c}}, Z. and 
	{Kent}, S. and {Knapp}, G.~R. and {Lin}, H. and {Loh}, Y.-S. and 
	{Lupton}, R.~H. and {Margon}, B. and {McKay}, T.~A. and {Meiksin}, A. and 
	{Munn}, J.~A. and {Pope}, A. and {Richmond}, M.~W. and {Schlegel}, D. and 
	{Schneider}, D.~P. and {Shimasaku}, K. and {Stoughton}, C. and 
	{Strauss}, M.~A. and {SubbaRao}, M. and {Szalay}, A.~S. and 
	{Szapudi}, I. and {Tucker}, D.~L. and {Yanny}, B. and {York}, D.~G.
	},
    title = "{Detection of the Baryon Acoustic Peak in the Large-Scale Correlation Function of SDSS Luminous Red Galaxies}",
  journal = {\apj},
   eprint = {astro-ph/0501171},
     year = 2005,
    month = nov,
   volume = 633,
    pages = {560-574},
      doi = {10.1086/466512},
   adsurl = {https://ui.adsabs.harvard.edu/abs/2005ApJ...633..560E}
}

@ARTICLE{Percival:2007yw,
   author = {{Percival}, W.~J. and {Cole}, S. and {Eisenstein}, D.~J. and 
	{Nichol}, R.~C. and {Peacock}, J.~A. and {Pope}, A.~C. and {Szalay}, A.~S.
	},
    title = "{Measuring the Baryon Acoustic Oscillation scale using the Sloan Digital Sky Survey and 2dF Galaxy Redshift Survey}",
  journal = {mnras},
archivePrefix = "arXiv",
   eprint = {0705.3323},
     year = 2007,
    month = nov,
   volume = 381,
    pages = {1053-1066},
      doi = {10.1111/j.1365-2966.2007.12268.x},
   adsurl = {https://ui.adsabs.harvard.edu/abs/2007MNRAS.381.1053P}
}

@ARTICLE{anderson2012clustering,
   author = {{Anderson}, L. and {Aubourg}, E. and {Bailey}, S. and {Bizyaev}, D. and 
	{Blanton}, M. and {Bolton}, A.~S. and {Brinkmann}, J. and {Brownstein}, J.~R. and 
	{Burden}, A. and {Cuesta}, A.~J. and {da Costa}, L.~A.~N. and 
	{Dawson}, K.~S. and {de Putter}, R. and {Eisenstein}, D.~J. and 
	{Gunn}, J.~E. and {Guo}, H. and {Hamilton}, J.-C. and {Harding}, P. and 
	{Ho}, S. and {Honscheid}, K. and {Kazin}, E. and {Kirkby}, D. and 
	{Kneib}, J.-P. and {Labatie}, A. and {Loomis}, C. and {Lupton}, R.~H. and 
	{Malanushenko}, E. and {Malanushenko}, V. and {Mandelbaum}, R. and 
	{Manera}, M. and {Maraston}, C. and {McBride}, C.~K. and {Mehta}, K.~T. and 
	{Mena}, O. and {Montesano}, F. and {Muna}, D. and {Nichol}, R.~C. and 
	{Nuza}, S.~E. and {Olmstead}, M.~D. and {Oravetz}, D. and {Padmanabhan}, N. and 
	{Palanque-Delabrouille}, N. and {Pan}, K. and {Parejko}, J. and 
	{P{\^a}ris}, I. and {Percival}, W.~J. and {Petitjean}, P. and 
	{Prada}, F. and {Reid}, B. and {Roe}, N.~A. and {Ross}, A.~J. and 
	{Ross}, N.~P. and {Samushia}, L. and {S{\'a}nchez}, A.~G. and 
	{Schlegel}, D.~J. and {Schneider}, D.~P. and {Sc{\'o}ccola}, C.~G. and 
	{Seo}, H.-J. and {Sheldon}, E.~S. and {Simmons}, A. and {Skibba}, R.~A. and 
	{Strauss}, M.~A. and {Swanson}, M.~E.~C. and {Thomas}, D. and 
	{Tinker}, J.~L. and {Tojeiro}, R. and {Maga{\~n}a}, M.~V. and 
	{Verde}, L. and {Wagner}, C. and {Wake}, D.~A. and {Weaver}, B.~A. and 
	{Weinberg}, D.~H. and {White}, M. and {Xu}, X. and {Y{\`e}che}, C. and 
	{Zehavi}, I. and {Zhao}, G.-B.},
    title = "{The clustering of galaxies in the SDSS-III Baryon Oscillation Spectroscopic Survey: baryon acoustic oscillations in the Data Release 9 spectroscopic galaxy sample}",
  journal = {mnras},
archivePrefix = "arXiv",
   eprint = {1203.6594},
     year = 2012,
    month = dec,
   volume = 427,
    pages = {3435-3467},
      doi = {10.1111/j.1365-2966.2012.22066.x},
   adsurl = {https://ui.adsabs.harvard.edu/abs/2012MNRAS.427.3435A}
}

@ARTICLE{alam2017clustering,
   author = {{Alam}, S. and {Ata}, M. and {Bailey}, S. and {Beutler}, F. and 
	{Bizyaev}, D. and {Blazek}, J.~A. and {Bolton}, A.~S. and {Brownstein}, J.~R. and 
	{Burden}, A. and {Chuang}, C.-H. and {Comparat}, J. and {Cuesta}, A.~J. and 
	{Dawson}, K.~S. and {Eisenstein}, D.~J. and {Escoffier}, S. and 
	{Gil-Mar{\'{\i}}n}, H. and {Grieb}, J.~N. and {Hand}, N. and 
	{Ho}, S. and {Kinemuchi}, K. and {Kirkby}, D. and {Kitaura}, F. and 
	{Malanushenko}, E. and {Malanushenko}, V. and {Maraston}, C. and 
	{McBride}, C.~K. and {Nichol}, R.~C. and {Olmstead}, M.~D. and 
	{Oravetz}, D. and {Padmanabhan}, N. and {Palanque-Delabrouille}, N. and 
	{Pan}, K. and {Pellejero-Ibanez}, M. and {Percival}, W.~J. and 
	{Petitjean}, P. and {Prada}, F. and {Price-Whelan}, A.~M. and 
	{Reid}, B.~A. and {Rodr{\'{\i}}guez-Torres}, S.~A. and {Roe}, N.~A. and 
	{Ross}, A.~J. and {Ross}, N.~P. and {Rossi}, G. and {Rubi{\~n}o-Mart{\'{\i}}n}, J.~A. and 
	{Saito}, S. and {Salazar-Albornoz}, S. and {Samushia}, L. and 
	{S{\'a}nchez}, A.~G. and {Satpathy}, S. and {Schlegel}, D.~J. and 
	{Schneider}, D.~P. and {Sc{\'o}ccola}, C.~G. and {Seo}, H.-J. and 
	{Sheldon}, E.~S. and {Simmons}, A. and {Slosar}, A. and {Strauss}, M.~A. and 
	{Swanson}, M.~E.~C. and {Thomas}, D. and {Tinker}, J.~L. and 
	{Tojeiro}, R. and {Maga{\~n}a}, M.~V. and {Vazquez}, J.~A. and 
	{Verde}, L. and {Wake}, D.~A. and {Wang}, Y. and {Weinberg}, D.~H. and 
	{White}, M. and {Wood-Vasey}, W.~M. and {Y{\`e}che}, C. and 
	{Zehavi}, I. and {Zhai}, Z. and {Zhao}, G.-B.},
    title = "{The clustering of galaxies in the completed SDSS-III Baryon Oscillation Spectroscopic Survey: cosmological analysis of the DR12 galaxy sample}",
  journal = {mnras},
archivePrefix = "arXiv",
   eprint = {1607.03155},
     year = 2017,
    month = sep,
   volume = 470,
    pages = {2617-2652},
      doi = {10.1093/mnras/stx721},
   adsurl = {https://ui.adsabs.harvard.edu/abs/2017MNRAS.470.2617A}
}

@ARTICLE{Kaiser,
       author = {{Kaiser}, Nick},
        title = "{Clustering in real space and in redshift space}",
      journal = {Monthly Notices of the Royal Astronomical Society},
         year = 1987,
        month = jul,
       volume = {227},
        pages = {1-21},
          doi = {10.1093/mnras/227.1.1},
       adsurl = {https://ui.adsabs.harvard.edu/abs/1987MNRAS.227....1K}
}

@ARTICLE{Ballinger,
       author = {{Ballinger}, W.~E. and {Peacock}, J.~A. and {Heavens}, A.~F.},
        title = "{Measuring the cosmological constant with redshift surveys}",
      journal = {Monthly Notices of the Royal Astronomical Society},
         year = 1996,
        month = oct,
       volume = {282},
        pages = {877},
          doi = {10.1093/mnras/282.3.877},
archivePrefix = {arXiv},
       eprint = {astro-ph/9605017},
 primaryClass = {astro-ph},
       adsurl = {https://ui.adsabs.harvard.edu/abs/1996MNRAS.282..877B}
}

@ARTICLE{Eisenstein_1998,
       author = {{Eisenstein}, Daniel J. and {Hu}, Wayne},
        title = "{Baryonic Features in the Matter Transfer Function}",
      journal = {The Astrophysical Journal},
         year = 1998,
        month = mar,
       volume = {496},
       number = {2},
        pages = {605-614},
          doi = {10.1086/305424},
archivePrefix = {arXiv},
       eprint = {astro-ph/9709112},
 primaryClass = {astro-ph},
       adsurl = {https://ui.adsabs.harvard.edu/abs/1998ApJ...496..605E}
}

@ARTICLE{Blake_2003,
       author = {{Blake}, Chris and {Glazebrook}, Karl},
        title = "{Probing Dark Energy Using Baryonic Oscillations in the Galaxy Power Spectrum as a Cosmological Ruler}",
      journal = {The Astrophysical Journal},
         year = 2003,
        month = sep,
       volume = {594},
       number = {2},
        pages = {665-673},
          doi = {10.1086/376983},
archivePrefix = {arXiv},
       eprint = {astro-ph/0301632},
 primaryClass = {astro-ph},
       adsurl = {https://ui.adsabs.harvard.edu/abs/2003ApJ...594..665B}
}

@ARTICLE{Seo_2003,
       author = {{Seo}, Hee-Jong and {Eisenstein}, Daniel J.},
        title = "{Probing Dark Energy with Baryonic Acoustic Oscillations from Future Large Galaxy Redshift Surveys}",
      journal = {The Astrophysical Journal},
         year = 2003,
        month = dec,
       volume = {598},
       number = {2},
        pages = {720-740},
          doi = {10.1086/379122},
archivePrefix = {arXiv},
       eprint = {astro-ph/0307460},
 primaryClass = {astro-ph},
       adsurl = {https://ui.adsabs.harvard.edu/abs/2003ApJ...598..720S}
}

@ARTICLE{2dFGRS,
       author = {{Colless}, Matthew and {Peterson}, Bruce A. and {Jackson}, Carole and {Peacock}, John A. and {Cole}, Shaun and {Norberg}, Peder and {Baldry}, Ivan K. and {Baugh}, Carlton M. and {Bland-Hawthorn}, Joss and {Bridges}, Terry and {Cannon}, Russell and {Collins}, Chris and {Couch}, Warrick and {Cross}, Nicholas and {Dalton}, Gavin and {De Propris}, Roberto and {Driver}, Simon P. and {Efstathiou}, George and {Ellis}, Richard S. and {Frenk}, Carlos S. and {Glazebrook}, Karl and {Lahav}, Ofer and {Lewis}, Ian and {Lumsden}, Stuart and {Maddox}, Steve and {Madgwick}, Darren and {Sutherland}, Will and {Taylor}, Keith},
        title = "{The 2dF Galaxy Redshift Survey: Final Data Release}",
      journal = {arXiv e-prints},
         year = 2003,
        month = jun,
          eid = {astro-ph/0306581},
        pages = {astro-ph/0306581},
          doi = {10.48550/arXiv.astro-ph/0306581},
archivePrefix = {arXiv},
       eprint = {astro-ph/0306581},
 primaryClass = {astro-ph},
       adsurl = {https://ui.adsabs.harvard.edu/abs/2003astro.ph..6581C}
}

@article{6dFGRS,
    author = {Beutler, Florian and Blake, Chris and Colless, Matthew and Jones, D. Heath and Staveley-Smith, Lister and Poole, Gregory B. and Campbell, Lachlan and Parker, Quentin and Saunders, Will and Watson, Fred},
    title = "{The 6dF Galaxy Survey: z≈ 0 measurements of the growth rate and σ8}",
    journal = {Monthly Notices of the Royal Astronomical Society},
    volume = {423},
    number = {4},
    pages = {3430-3444},
    year = {2012},
    month = {07},
    issn = {0035-8711},
    doi = {10.1111/j.1365-2966.2012.21136.x},
    url = {https://doi.org/10.1111/j.1365-2966.2012.21136.x},
    eprint = {https://academic.oup.com/mnras/article-pdf/423/4/3430/4903419/mnras0423-3430.pdf},
}

@ARTICLE{WiggleZ2011B,
       author = {{Blake}, Chris and {Glazebrook}, Karl and {Davis}, Tamara M. and {Brough}, Sarah and {Colless}, Matthew and {Contreras}, Carlos and {Couch}, Warrick and {Croom}, Scott and {Drinkwater}, Michael J. and {Forster}, Karl and {Gilbank}, David and {Gladders}, Mike and {Jelliffe}, Ben and {Jurek}, Russell J. and {Li}, I. -Hui and {Madore}, Barry and {Martin}, D. Christopher and {Pimbblet}, Kevin and {Poole}, Gregory B. and {Pracy}, Michael and {Sharp}, Rob and {Wisnioski}, Emily and {Woods}, David and {Wyder}, Ted K. and {Yee}, H.~K.~C.},
        title = "{The WiggleZ Dark Energy Survey: measuring the cosmic expansion history using the Alcock-Paczynski test and distant supernovae}",
      journal = {Monthly Notices of the Royal Astronomical Society},
         year = 2011,
        month = dec,
       volume = {418},
       number = {3},
        pages = {1725-1735},
          doi = {10.1111/j.1365-2966.2011.19606.x},
archivePrefix = {arXiv},
       eprint = {1108.2637},
 primaryClass = {astro-ph.CO},
       adsurl = {https://ui.adsabs.harvard.edu/abs/2011MNRAS.418.1725B}
}

@ARTICLE{WiggleZ2011c,
       author = {{Blake}, Chris and {Brough}, Sarah and {Colless}, Matthew and {Contreras}, Carlos and {Couch}, Warrick and {Croom}, Scott and {Davis}, Tamara and {Drinkwater}, Michael J. and {Forster}, Karl and {Gilbank}, David and {Gladders}, Mike and {Glazebrook}, Karl and {Jelliffe}, Ben and {Jurek}, Russell J. and {Li}, I. -Hui and {Madore}, Barry and {Martin}, D. Christopher and {Pimbblet}, Kevin and {Poole}, Gregory B. and {Pracy}, Michael and {Sharp}, Rob and {Wisnioski}, Emily and {Woods}, David and {Wyder}, Ted K. and {Yee}, H.~K.~C.},
        title = "{The WiggleZ Dark Energy Survey: the growth rate of cosmic structure since redshift z=0.9}",
      journal = {Monthly Notices of the Royal Astronomical Society},
         year = 2011,
        month = aug,
       volume = {415},
       number = {3},
        pages = {2876-2891},
          doi = {10.1111/j.1365-2966.2011.18903.x},
archivePrefix = {arXiv},
       eprint = {1104.2948},
 primaryClass = {astro-ph.CO},
       adsurl = {https://ui.adsabs.harvard.edu/abs/2011MNRAS.415.2876B}
}

@ARTICLE{SDSS_York,
       author = {{York}, Donald G. and {Adelman}, J. and {Anderson}, John E., Jr. and {Anderson}, Scott F. and {Annis}, James and {Bahcall}, Neta A. and {Bakken}, J.~A. and {Barkhouser}, Robert and {Bastian}, Steven and {Berman}, Eileen and {Boroski}, William N. and {Bracker}, Steve and {Briegel}, Charlie and {Briggs}, John W. and {Brinkmann}, J. and {Brunner}, Robert and {Burles}, Scott and {Carey}, Larry and {Carr}, Michael A. and {Castander}, Francisco J. and {Chen}, Bing and {Colestock}, Patrick L. and {Connolly}, A.~J. and {Crocker}, J.~H. and {Csabai}, Istv{\'a}n and {Czarapata}, Paul C. and {Davis}, John Eric and {Doi}, Mamoru and {Dombeck}, Tom and {Eisenstein}, Daniel and {Ellman}, Nancy and {Elms}, Brian R. and {Evans}, Michael L. and {Fan}, Xiaohui and {Federwitz}, Glenn R. and {Fiscelli}, Larry and {Friedman}, Scott and {Frieman}, Joshua A. and {Fukugita}, Masataka and {Gillespie}, Bruce and {Gunn}, James E. and {Gurbani}, Vijay K. and {de Haas}, Ernst and {Haldeman}, Merle and {Harris}, Frederick H. and {Hayes}, J. and {Heckman}, Timothy M. and {Hennessy}, G.~S. and {Hindsley}, Robert B. and {Holm}, Scott and {Holmgren}, Donald J. and {Huang}, Chi-hao and {Hull}, Charles and {Husby}, Don and {Ichikawa}, Shin-Ichi and {Ichikawa}, Takashi and {Ivezi{\'c}}, {\v{Z}}eljko and {Kent}, Stephen and {Kim}, Rita S.~J. and {Kinney}, E. and {Klaene}, Mark and {Kleinman}, A.~N. and {Kleinman}, S. and {Knapp}, G.~R. and {Korienek}, John and {Kron}, Richard G. and {Kunszt}, Peter Z. and {Lamb}, D.~Q. and {Lee}, B. and {Leger}, R. French and {Limmongkol}, Siriluk and {Lindenmeyer}, Carl and {Long}, Daniel C. and {Loomis}, Craig and {Loveday}, Jon and {Lucinio}, Rich and {Lupton}, Robert H. and {MacKinnon}, Bryan and {Mannery}, Edward J. and {Mantsch}, P.~M. and {Margon}, Bruce and {McGehee}, Peregrine and {McKay}, Timothy A. and {Meiksin}, Avery and {Merelli}, Aronne and {Monet}, David G. and {Munn}, Jeffrey A. and {Narayanan}, Vijay K. and {Nash}, Thomas and {Neilsen}, Eric and {Neswold}, Rich and {Newberg}, Heidi Jo and {Nichol}, R.~C. and {Nicinski}, Tom and {Nonino}, Mario and {Okada}, Norio and {Okamura}, Sadanori and {Ostriker}, Jeremiah P. and {Owen}, Russell and {Pauls}, A. George and {Peoples}, John and {Peterson}, R.~L. and {Petravick}, Donald and {Pier}, Jeffrey R. and {Pope}, Adrian and {Pordes}, Ruth and {Prosapio}, Angela and {Rechenmacher}, Ron and {Quinn}, Thomas R. and {Richards}, Gordon T. and {Richmond}, Michael W. and {Rivetta}, Claudio H. and {Rockosi}, Constance M. and {Ruthmansdorfer}, Kurt and {Sandford}, Dale and {Schlegel}, David J. and {Schneider}, Donald P. and {Sekiguchi}, Maki and {Sergey}, Gary and {Shimasaku}, Kazuhiro and {Siegmund}, Walter A. and {Smee}, Stephen and {Smith}, J. Allyn and {Snedden}, S. and {Stone}, R. and {Stoughton}, Chris and {Strauss}, Michael A. and {Stubbs}, Christopher and {SubbaRao}, Mark and {Szalay}, Alexander S. and {Szapudi}, Istvan and {Szokoly}, Gyula P. and {Thakar}, Anirudda R. and {Tremonti}, Christy and {Tucker}, Douglas L. and {Uomoto}, Alan and {Vanden Berk}, Dan and {Vogeley}, Michael S. and {Waddell}, Patrick and {Wang}, Shu-i. and {Watanabe}, Masaru and {Weinberg}, David H. and {Yanny}, Brian and {Yasuda}, Naoki and {SDSS Collaboration}},
        title = "{The Sloan Digital Sky Survey: Technical Summary}",
      journal = {The Astronomical Journal},
         year = 2000,
        month = sep,
       volume = {120},
       number = {3},
        pages = {1579-1587},
          doi = {10.1086/301513},
archivePrefix = {arXiv},
       eprint = {astro-ph/0006396},
 primaryClass = {astro-ph},
       adsurl = {https://ui.adsabs.harvard.edu/abs/2000AJ....120.1579Y}
}

@ARTICLE{sanchez2012clustering,
       author = {{S{\'a}nchez}, Ariel G. and {Sc{\'o}ccola}, C.~G. and {Ross}, A.~J. and
         {Percival}, W. and {Manera}, M. and {Montesano}, F. and {Mazzalay}, X. and
         {Cuesta}, A.~J. and {Eisenstein}, D.~J. and {Kazin}, E. and
         {McBride}, C.~K. and {Mehta}, K. and {Montero-Dorta}, A.~D. and
         {Padmanabhan}, N. and {Prada}, F. and {Rubi{\~n}o-Mart{\'\i}n}, J.~A. and
         {Tojeiro}, R. and {Xu}, X. and {Maga{\~n}a}, M. Vargas and
         {Aubourg}, E. and {Bahcall}, N.~A. and {Bailey}, S. and {Bizyaev}, D. and
         {Bolton}, A.~S. and {Brewington}, H. and {Brinkmann}, J. and
         {Brownstein}, J.~R. and {Gott}, J. Richard and {Hamilton}, J.~C. and
         {Ho}, S. and {Honscheid}, K. and {Labatie}, A. and {Malanushenko}, E. and
         {Malanushenko}, V. and {Maraston}, C. and {Muna}, D. and
         {Nichol}, R.~C. and {Oravetz}, D. and {Pan}, K. and {Ross}, N.~P. and
         {Roe}, N.~A. and {Reid}, B.~A. and {Schlegel}, D.~J. and {Shelden}, A. and
         {Schneider}, D.~P. and {Simmons}, A. and {Skibba}, R. and
         {Snedden}, S. and {Thomas}, D. and {Tinker}, J. and {Wake}, D.~A. and
         {Weaver}, B.~A. and {Weinberg}, David H. and {White}, Martin and
         {Zehavi}, I. and {Zhao}, G.},
        title = "{The clustering of galaxies in the SDSS-III Baryon Oscillation Spectroscopic Survey: cosmological implications of the large-scale two-point correlation function}",
      journal = {Monthly Notices of the Royal Astronomical Society},
         year = 2012,
        month = sep,
       volume = {425},
       number = {1},
        pages = {415-437},
          doi = {10.1111/j.1365-2966.2012.21502.x},
archivePrefix = {arXiv},
       eprint = {1203.6616},
 primaryClass = {astro-ph.CO},
       adsurl = {https://ui.adsabs.harvard.edu/abs/2012MNRAS.425..415S}
}

@ARTICLE{sanchez2013clustering,
       author = {{S{\'a}nchez}, Ariel G. and {Kazin}, Eyal A. and {Beutler}, Florian and
         {Chuang}, Chia-Hsun and {Cuesta}, Antonio J. and
         {Eisenstein}, Daniel J. and {Manera}, Marc and {Montesano}, Francesco and
         {Nichol}, Robert C. and {Padmanabhan}, Nikhil and {Percival}, Will and
         {Prada}, Francisco and {Ross}, Ashley J. and {Schlegel}, David J. and
         {Tinker}, Jeremy and {Tojeiro}, Rita and {Weinberg}, David H. and
         {Xu}, Xiaoying and {Brinkmann}, J. and {Brownstein}, Joel R. and
         {Schneider}, Donald P. and {Thomas}, Daniel},
        title = "{The clustering of galaxies in the SDSS-III Baryon Oscillation Spectroscopic Survey: cosmological constraints from the full shape of the clustering wedges}",
      journal = {Monthly Notices of the Royal Astronomical Society},
         year = 2013,
        month = aug,
       volume = {433},
       number = {2},
        pages = {1202-1222},
          doi = {10.1093/mnras/stt799},
archivePrefix = {arXiv},
       eprint = {1303.4396},
 primaryClass = {astro-ph.CO},
       adsurl = {https://ui.adsabs.harvard.edu/abs/2013MNRAS.433.1202S}
}

@ARTICLE{anderson2014clustering,
       author = {{Anderson}, Lauren and {Aubourg}, {\'E}ric and {Bailey}, Stephen and
         {Beutler}, Florian and {Bhardwaj}, Vaishali and {Blanton}, Michael and
         {Bolton}, Adam S. and {Brinkmann}, J. and {Brownstein}, Joel R. and
         {Burden}, Angela and {Chuang}, Chia-Hsun and {Cuesta}, Antonio J. and
         {Dawson}, Kyle S. and {Eisenstein}, Daniel J. and
         {Escoffier}, Stephanie and {Gunn}, James E. and {Guo}, Hong and
         {Ho}, Shirley and {Honscheid}, Klaus and {Howlett}, Cullan and
         {Kirkby}, David and {Lupton}, Robert H. and {Manera}, Marc and
         {Maraston}, Claudia and {McBride}, Cameron K. and {Mena}, Olga and
         {Montesano}, Francesco and {Nichol}, Robert C. and
         {Nuza}, Sebasti{\'a}n E. and {Olmstead}, Matthew D. and
         {Padmanabhan}, Nikhil and {Palanque-Delabrouille}, Nathalie and
         {Parejko}, John and {Percival}, Will J. and {Petitjean}, Patrick and
         {Prada}, Francisco and {Price-Whelan}, Adrian M. and {Reid}, Beth and
         {Roe}, Natalie A. and {Ross}, Ashley J. and {Ross}, Nicholas P. and
         {Sabiu}, Cristiano G. and {Saito}, Shun and {Samushia}, Lado and
         {S{\'a}nchez}, Ariel G. and {Schlegel}, David J. and
         {Schneider}, Donald P. and {Scoccola}, Claudia G. and {Seo}, Hee-Jong and
         {Skibba}, Ramin A. and {Strauss}, Michael A. and
         {Swanson}, Molly E.~C. and {Thomas}, Daniel and {Tinker}, Jeremy L. and
         {Tojeiro}, Rita and {Maga{\~n}a}, Mariana Vargas and {Verde}, Licia and
         {Wake}, David A. and {Weaver}, Benjamin A. and {Weinberg}, David H. and
         {White}, Martin and {Xu}, Xiaoying and {Y{\`e}che}, Christophe and
         {Zehavi}, Idit and {Zhao}, Gong-Bo},
        title = "{The clustering of galaxies in the SDSS-III Baryon Oscillation Spectroscopic Survey: baryon acoustic oscillations in the Data Releases 10 and 11 Galaxy samples}",
      journal = {Monthly Notices of the Royal Astronomical Society},
         year = 2014,
        month = jun,
       volume = {441},
       number = {1},
        pages = {24-62},
          doi = {10.1093/mnras/stu523},
archivePrefix = {arXiv},
       eprint = {1312.4877},
 primaryClass = {astro-ph.CO},
       adsurl = {https://ui.adsabs.harvard.edu/abs/2014MNRAS.441...24A}
}

@ARTICLE{samushia2014clustering,
       author = {{Samushia}, Lado and {Reid}, Beth A. and {White}, Martin and
         {Percival}, Will J. and {Cuesta}, Antonio J. and {Zhao}, Gong-Bo and
         {Ross}, Ashley J. and {Manera}, Marc and {Aubourg}, {\'E}ric and
         {Beutler}, Florian and {Brinkmann}, Jon and {Brownstein}, Joel R. and
         {Dawson}, Kyle S. and {Eisenstein}, Daniel J. and {Ho}, Shirley and
         {Honscheid}, Klaus and {Maraston}, Claudia and {Montesano}, Francesco and
         {Nichol}, Robert C. and {Roe}, Natalie A. and {Ross}, Nicholas P. and
         {S{\'a}nchez}, Ariel G. and {Schlegel}, David J. and
         {Schneider}, Donald P. and {Streblyanska}, Alina and {Thomas}, Daniel and
         {Tinker}, Jeremy L. and {Wake}, David A. and {Weaver}, Benjamin A. and
         {Zehavi}, Idit},
        title = "{The clustering of galaxies in the SDSS-III Baryon Oscillation Spectroscopic Survey: measuring growth rate and geometry with anisotropic clustering}",
      journal = {Monthly Notices of the Royal Astronomical Society},
         year = 2014,
        month = apr,
       volume = {439},
       number = {4},
        pages = {3504-3519},
          doi = {10.1093/mnras/stu197},
archivePrefix = {arXiv},
       eprint = {1312.4899},
 primaryClass = {astro-ph.CO},
       adsurl = {https://ui.adsabs.harvard.edu/abs/2014MNRAS.439.3504S}
}

@ARTICLE{ross2015clustering,
       author = {{Ross}, Ashley J. and {Samushia}, Lado and {Howlett}, Cullan and
         {Percival}, Will J. and {Burden}, Angela and {Manera}, Marc},
        title = "{The clustering of the SDSS DR7 main Galaxy sample - I. A 4 per cent distance measure at z = 0.15}",
      journal = {Monthly Notices of the Royal Astronomical Society},
         year = 2015,
        month = may,
       volume = {449},
       number = {1},
        pages = {835-847},
          doi = {10.1093/mnras/stv154},
archivePrefix = {arXiv},
       eprint = {1409.3242},
 primaryClass = {astro-ph.CO},
       adsurl = {https://ui.adsabs.harvard.edu/abs/2015MNRAS.449..835R}
}

@ARTICLE{beutler2016clustering,
       author = {{Beutler}, Florian and {Seo}, Hee-Jong and {Ross}, Ashley J. and
         {McDonald}, Patrick and {Saito}, Shun and {Bolton}, Adam S. and
         {Brownstein}, Joel R. and {Chuang}, Chia-Hsun and {Cuesta}, Antonio J. and
         {Eisenstein}, Daniel J. and {Font-Ribera}, Andreu and
         {Grieb}, Jan Niklas and {Hand}, Nick and {Kitaura}, Francisco-Shu and
         {Modi}, Chirag and {Nichol}, Robert C. and {Percival}, Will J. and
         {Prada}, Francisco and {Rodriguez-Torres}, Sergio and
         {Roe}, Natalie A. and {Ross}, Nicholas P. and
         {Salazar-Albornoz}, Salvador and {S{\'a}nchez}, Ariel G. and
         {Schneider}, Donald P. and {Slosar}, An{\v{z}}e and {Tinker}, Jeremy and
         {Tojeiro}, Rita and {Vargas-Maga{\~n}a}, Mariana and {Vazquez}, Jose A.},
        title = "{The clustering of galaxies in the completed SDSS-III Baryon Oscillation Spectroscopic Survey: baryon acoustic oscillations in the Fourier space}",
      journal = {Monthly Notices of the Royal Astronomical Society},
         year = 2017,
        month = jan,
       volume = {464},
       number = {3},
        pages = {3409-3430},
          doi = {10.1093/mnras/stw2373},
archivePrefix = {arXiv},
       eprint = {1607.03149},
 primaryClass = {astro-ph.CO},
       adsurl = {https://ui.adsabs.harvard.edu/abs/2017MNRAS.464.3409B}
}

@ARTICLE{sanchez2016clustering,
       author = {{S{\'a}nchez}, Ariel G. and {Grieb}, Jan Niklas and
         {Salazar-Albornoz}, Salvador and {Alam}, Shadab and {Beutler}, Florian and
         {Ross}, Ashley J. and {Brownstein}, Joel R. and {Chuang}, Chia-Hsun and
         {Cuesta}, Antonio J. and {Eisenstein}, Daniel J. and
         {Kitaura}, Francisco-Shu and {Percival}, Will J. and
         {Prada}, Francisco and {Rodr{\'\i}guez-Torres}, Sergio and
         {Seo}, Hee-Jong and {Tinker}, Jeremy and {Tojeiro}, Rita and
         {Vargas-Maga{\~n}a}, Mariana and {Vazquez}, Jose A. and {Zhao}, Gong-Bo},
        title = "{The clustering of galaxies in the completed SDSS-III Baryon Oscillation Spectroscopic Survey: combining correlated Gaussian posterior distributions}",
      journal = {Monthly Notices of the Royal Astronomical Society},
         year = 2017,
        month = jan,
       volume = {464},
       number = {2},
        pages = {1493-1501},
          doi = {10.1093/mnras/stw2495},
archivePrefix = {arXiv},
       eprint = {1607.03146},
 primaryClass = {astro-ph.CO},
       adsurl = {https://ui.adsabs.harvard.edu/abs/2017MNRAS.464.1493S}
}

@ARTICLE{chuang2017clustering,
       author = {{Chuang}, Chia-Hsun and {Kitaura}, Francisco-Shu and {Liang}, Yu and
         {Font-Ribera}, Andreu and {Zhao}, Cheng and {McDonald}, Patrick and
         {Tao}, Charling},
        title = "{Linear redshift space distortions for cosmic voids based on galaxies in redshift space}",
      journal = {Physical Review D },
         year = 2017,
        month = mar,
       volume = {95},
       number = {6},
          eid = {063528},
        pages = {063528},
          doi = {10.1103/PhysRevD.95.063528},
archivePrefix = {arXiv},
       eprint = {1605.05352},
 primaryClass = {astro-ph.CO},
       adsurl = {https://ui.adsabs.harvard.edu/abs/2017PhRvD..95f3528C}
}

@ARTICLE{Sabiu2016A&A,
       author = {{Sabiu}, Cristiano G. and {Mota}, David F. and {Llinares}, Claudio and {Park}, Changbom},
        title = "{Probing scalar tensor theories for gravity in redshift space}",
      journal = {Astronomy and Astrophysics },
         year = 2016,
        month = jul,
       volume = {592},
          eid = {A38},
        pages = {A38},
          doi = {10.1051/0004-6361/201527776},
archivePrefix = {arXiv},
       eprint = {1603.05750},
 primaryClass = {astro-ph.CO},
       adsurl = {https://ui.adsabs.harvard.edu/abs/2016A&A...592A..38S}
}

@ARTICLE{Slepian_2017,
       author = {{Slepian}, Zachary and {Eisenstein}, Daniel J. and
         {Brownstein}, Joel R. and {Chuang}, Chia-Hsun and
         {Gil-Mar{\'\i}n}, H{\'e}ctor and {Ho}, Shirley and
         {Kitaura}, Francisco-Shu and {Percival}, Will J. and {Ross}, Ashley J. and
         {Rossi}, Graziano and {Seo}, Hee-Jong and {Slosar}, An{\v{z}}e and
         {Vargas-Maga{\~n}a}, Mariana},
        title = "{Detection of baryon acoustic oscillation features in the large-scale three-point correlation function of SDSS BOSS DR12 CMASS galaxies}",
      journal = {Monthly Notices of the Royal Astronomical Society},
         year = 2017,
        month = aug,
       volume = {469},
       number = {2},
        pages = {1738-1751},
          doi = {10.1093/mnras/stx488},
archivePrefix = {arXiv},
       eprint = {1607.06097},
 primaryClass = {astro-ph.CO},
       adsurl = {https://ui.adsabs.harvard.edu/abs/2017MNRAS.469.1738S}
}

@ARTICLE{Sabiu_2019,
       author = {{Sabiu}, Cristiano G. and {Hoyle}, Ben and {Kim}, Juhan and
         {Li}, Xiao-Dong},
        title = "{Graph Database Solution for Higher-order Spatial Statistics in the Era of Big Data}",
      journal = {The Astrophysical Journals},
         year = 2019,
        month = jun,
       volume = {242},
       number = {2},
          eid = {29},
        pages = {29},
          doi = {10.3847/1538-4365/ab22b5},
archivePrefix = {arXiv},
       eprint = {1901.00296},
 primaryClass = {astro-ph.CO},
       adsurl = {https://ui.adsabs.harvard.edu/abs/2019ApJS..242...29S}
}

@ARTICLE{ryden1995measuring,
   author = {{Ryden}, B.~S.},
    title = "{Measuring Q 0 from the Distortion of Voids in Redshift Space}",
  journal = {The Astrophysical Journal},
   eprint = {astro-ph/9506028},
     year = 1995,
    month = oct,
   volume = 452,
    pages = {25},
      doi = {10.1086/176277},
   adsurl = {https://ui.adsabs.harvard.edu/abs/1995ApJ...452...25R}
}

@ARTICLE{lavaux2012precision,
   author = {{Lavaux}, G. and {Wandelt}, B.~D.},
    title = "{Precision Cosmography with Stacked Voids}",
  journal = {The Astrophysical Journal},
archivePrefix = "arXiv",
   eprint = {1110.0345},
 primaryClass = "astro-ph.CO",
     year = 2012,
    month = aug,
   volume = 754,
      eid = {109},
    pages = {109},
      doi = {10.1088/0004-637X/754/2/109},
   adsurl = {https://ui.adsabs.harvard.edu/abs/2012ApJ...754..109L}
}

@ARTICLE{Ravanbakhsh17,
   author = {{Ravanbakhsh}, S. and {Oliva}, J. and {Fromenteau}, S. and {Price}, L.~C. and 
	{Ho}, S. and {Schneider}, J. and {Poczos}, B.},
    title = "{Estimating Cosmological Parameters from the Dark Matter Distribution}",
  journal = {arXiv e-prints},
archivePrefix = "arXiv",
   eprint = {1711.02033},
     year = 2017,
    month = nov,
   adsurl = {https://ui.adsabs.harvard.edu/abs/2017arXiv171102033R}
}

@ARTICLE{Mathuriya18,
   author = {{Mathuriya}, A. and {Bard}, D. and {Mendygral}, P. and {Meadows}, L. and 
	{Arnemann}, J. and {Shao}, L. and {He}, S. and {Karna}, T. and 
	{Moise}, D. and {Pennycook}, S.~J. and {Maschoff}, K. and {Sewall}, J. and 
	{Kumar}, N. and {Ho}, S. and {Ringenburg}, M. and {Prabhat} and 
	{Lee}, V.},
    title = "{CosmoFlow: Using Deep Learning to Learn the Universe at Scale}",
  journal = {arXiv e-prints},
archivePrefix = "arXiv",
   eprint = {1808.04728},
     year = 2018,
    month = aug,
   adsurl = {https://ui.adsabs.harvard.edu/abs/2018arXiv180804728M}
}

@ARTICLE{Beisbart:2000ja,
       author = {{Beisbart}, Claus and {Kerscher}, Martin},
        title = "{Luminosity- and Morphology-dependent Clustering of Galaxies}",
      journal = {The Astrophysical Journal},
         year = 2000,
        month = dec,
       volume = {545},
       number = {1},
        pages = {6-25},
          doi = {10.1086/317788},
archivePrefix = {arXiv},
       eprint = {astro-ph/0003358},
 primaryClass = {astro-ph},
       adsurl = {https://ui.adsabs.harvard.edu/abs/2000ApJ...545....6B}
}

@INBOOK{Beisbart2002,
       author = {{Beisbart}, Claus and {Kerscher}, Martin and {Mecke}, Klaus},
    booktitle = {Morphology of Condensed Matter},
         year = 2002,
       editor = {{Mecke}, K. and {Stoyan}, D.},
       volume = {600},
        pages = {358-390},
       adsurl = {https://ui.adsabs.harvard.edu/abs/2002LNP...600..358B}
}

@ARTICLE{Gottl2002,
       author = {{Gottl{\"o}ber}, S. and {Kerscher}, M. and {Kravtsov}, A.~V. and
         {Faltenbacher}, A. and {Klypin}, A. and {M{\"u}ller}, V.},
        title = "{Spatial distribution of galactic halos and their merger histories}",
      journal = {Astronomy and Astrophysics },
         year = 2002,
        month = jun,
       volume = {387},
        pages = {778-787},
          doi = {10.1051/0004-6361:20020339},
archivePrefix = {arXiv},
       eprint = {astro-ph/0203148},
 primaryClass = {astro-ph},
       adsurl = {https://ui.adsabs.harvard.edu/abs/2002A&A...387..778G}
}

@ARTICLE{Sheth:2004vb,
       author = {{Sheth}, Ravi K. and {Tormen}, Giuseppe},
        title = "{On the environmental dependence of halo formation}",
      journal = {Monthly Notices of the Royal Astronomical Society},
         year = 2004,
        month = jun,
       volume = {350},
       number = {4},
        pages = {1385-1390},
          doi = {10.1111/j.1365-2966.2004.07733.x},
archivePrefix = {arXiv},
       eprint = {astro-ph/0402237},
 primaryClass = {astro-ph},
       adsurl = {https://ui.adsabs.harvard.edu/abs/2004MNRAS.350.1385S}
}

@ARTICLE{Sheth:2005aj,
       author = {{Sheth}, Ravi K. and {Connolly}, Andrew J. and {Skibba}, Ramin},
        title = "{Marked correlations in galaxy formation models}",
         year = "2005",
archivePrefix = "",
       eprint = "astro-ph/0511773",
}

@ARTICLE{Skibba2006,
       author = {{Skibba}, Ramin and {Sheth}, Ravi K. and {Connolly}, Andrew J. and
         {Scranton}, Ryan},
        title = "{The luminosity-weighted or `marked' correlation function}",
      journal = {Monthly Notices of the Royal Astronomical Society},
         year = 2006,
        month = jun,
       volume = {369},
       number = {1},
        pages = {68-76},
          doi = {10.1111/j.1365-2966.2006.10196.x},
archivePrefix = {arXiv},
       eprint = {astro-ph/0512463},
 primaryClass = {astro-ph},
       adsurl = {https://ui.adsabs.harvard.edu/abs/2006MNRAS.369...68S}
}

@ARTICLE{White_2009,
       author = {{White}, Martin and {Padmanabhan}, Nikhil},
        title = "{Breaking halo occupation degeneracies with marked statistics}",
      journal = {Monthly Notices of the Royal Astronomical Society},
         year = 2009,
        month = jun,
       volume = {395},
       number = {4},
        pages = {2381-2384},
          doi = {10.1111/j.1365-2966.2009.14732.x},
archivePrefix = {arXiv},
       eprint = {0812.4288},
 primaryClass = {astro-ph},
       adsurl = {https://ui.adsabs.harvard.edu/abs/2009MNRAS.395.2381W}
}

@ARTICLE{White2016,
       author = {{White}, Martin},
        title = "{A marked correlation function for constraining modified gravity models}",
      journal = "JCAP",
         year = 2016,
        month = nov,
       volume = {2016},
       number = {11},
          eid = {057},
        pages = {057},
          doi = {10.1088/1475-7516/2016/11/057},
archivePrefix = {arXiv},
       eprint = {1609.08632},
 primaryClass = {astro-ph.CO},
       adsurl = {https://ui.adsabs.harvard.edu/abs/2016JCAP...11..057W}
}

@ARTICLE{Satpathy:2019nvo,
       author = {{Satpathy}, Siddharth and {A C Croft}, Rupert and {Ho}, Shirley and
         {Li}, Baojiu},
        title = "{Measurement of marked correlation functions in SDSS-III Baryon Oscillation Spectroscopic Survey using LOWZ galaxies in Data Release 12}",
      journal = {Monthly Notices of the Royal Astronomical Society},
         year = 2019,
        month = apr,
       volume = {484},
       number = {2},
        pages = {2148-2165},
          doi = {10.1093/mnras/stz009},
archivePrefix = {arXiv},
       eprint = {1901.01447},
 primaryClass = {astro-ph.CO},
       adsurl = {https://ui.adsabs.harvard.edu/abs/2019MNRAS.484.2148S}
}

@article{massara2020,
  title={Using the marked power spectrum to detect the signature of neutrinos in large-scale structure},
  author={Massara, Elena and Villaescusa-Navarro, Francisco and Ho, Shirley and Dalal, Neal and Spergel, David N},
  journal={Physical Review Letters},
  volume={126},
  number={1},
  pages={011301},
  year={2021},
  publisher={APS}
}

@article{Philcox2020,
author = "Philcox, O. H. E. and Massara, E. and Spergel, D. N.",
    archivePrefix = "arXiv",
    eprint = "2006.10055",
    year = "2020",
    title = "{What does the Marked Power Spectrum Measure? Insights from Perturbation Theory}"
}

@article{pan2020cosmological,
  title={Cosmological parameter estimation from large-scale structure deep learning},
  author={Pan, ShuYang and Liu, MiaoXin and Forero-Romero, Jaime and Sabiu, Cristiano G and Li, ZhiGang and Miao, HaiTao and Li, Xiao-Dong},
  journal={Science China Physics, Mechanics \& Astronomy},
  volume={63},
  number={11},
  pages={110412},
  year={2020},
  publisher={Springer}
}

@article{sobol1967distribution,
  title={The distribution of points in a cube and the accurate evaluation of integrals (in Russian) Zh},
  author={Sobol, I},
  journal={Vychisl. Mat. i Mater. Phys},
  volume={7},
  pages={784--802},
  year={1967}
}

\end{document}